\documentclass[twocolumn]{aastex63}
\usepackage[varg]{txfonts}
\usepackage{natbib}
\usepackage{multirow}
\usepackage{hyperref}
\hypersetup{colorlinks=true, citecolor=blue}
\usepackage{xcolor}

\bibpunct{(}{)}{;}{a}{}{,}

\newcommand{\figref}[1]{Fig.\,\ref{#1}}
\newcommand{\tabref}[1]{Tab.\,\ref{#1}}

\newcommand{\tentothe}[1]{\,$\times$\,10$^{#1}$}
\newcommand{\Rsun}{\,R$_{\odot}$}
\newcommand{\Msun}{\,M$_{\odot}$}

\newcommand{\Halpha}{H${\alpha}$}
\newcommand{\citel}[1]{\citeauthor{#1}\,\citeyear{#1}}

\newcommand{\NH}{\,cm$^{-2}$}
\newcommand{\kmpersec}{\,km\,s$^{-1}$}
\newcommand{\p}{$\pm$}
\newcommand{\kelvin}{\,K}
\newcommand{\auperkpc}{\,au\,kpc$^{-1}$}
\newcommand{\Rsunperkpc}{\,\Rsun\,kpc$^{-1}$}
\newcommand{\brgamma}{Br$\gamma$}
\hypersetup{colorlinks=true, citecolor=blue}
\newcommand{\ttentothe}[1]{$\times$10$^{#1}$}

\newcommand{\Rstar}{R$_*$}
\newcommand{\Tstar}{T$_*$}
\newcommand{\Mstar}{M$_*$}
\newcommand{\Rrim}{R$_{rim}$}
\newcommand{\Hrim}{H$_{rim}$}
\newcommand{\Trim}{T$_{rim}$}
\newcommand{\Tin}{T$_{in}$}
\newcommand{\Rout}{R$_{out}$}

\begin{document}
\title{Optical and infrared study of the obscured B[e] supergiant high-mass X-ray binary IGR J16318-4848\thanks{Based on observations collected at the European Organisation for Astronomical Research in the Southern Hemisphere under ESO programme 089.D-0056.}}
\author{Francis Fortin} 
\affiliation{Universit\'e de Paris,
CNRS, Astroparticule et Cosmologie, F-75013 Paris, France}
\affiliation{AIM, CEA, CNRS, Universit\'e de Paris, Universit\'e Paris-Saclay, F-91191 Gif-sur-Yvette, France}

\author{Sylvain Chaty} 
\affiliation{Universit\'e de Paris,
CNRS, Astroparticule et Cosmologie, F-75013 Paris, France}
\affiliation{AIM, CEA, CNRS, Universit\'e de Paris, Universit\'e Paris-Saclay, F-91191 Gif-sur-Yvette, France}

\author{Andreas Sander} 
\affiliation{Armagh Observatory and Planetarium, College Hill, Armagh BT61 9DG, Northern Ireland}

\shorttitle{Optical and infrared study of IGR J16318-4848}
\shortauthors{F.\,Fortin et al.}

\correspondingauthor{Francis Fortin}

\begin{abstract}
The supergiant High Mass X-ray Binary IGR J16318-4848 was the first source detected by the \textit{INTEGRAL} satellite in 2003 and distinguishes itself by its high intrinsic absorption and B[e] phenomenon. It is the perfect candidate to study both binary interaction and the environment of supergiant B[e] stars.
This study targets the local properties of IGR J16318-4848. We aim to clarify the geometry of this system, and distinguish different key emitting regions in the binary.
We provide optical to near-infrared spectra from VLT/X-Shooter and analyse both fine structures of the lines and the broadband spectral energy distribution by adding archival mid-infrared \textit{Spitzer} and \textit{Herschel} data. We also performed a stellar atmosphere and wind modeling of the optical to near-infrared spectrum using the PoWR code.
We determine the contribution of the irradiated inner edge of the dusty circumbinary disk, derive the velocity of an equatorial stellar wind, and suggest the compact object orbits within the cavity between the star and the disk. We report on flat-topped lines originating from a spherically symetric disk wind, along with the first detection of what is likely the polar component of the stellar wind. Stellar atmosphere and wind modeling shows that the central star may have a helium-enhanced atmosphere, likely because of its intense wind sheding part of its hydrogen envelope. Finally we compare the properties of IGR J16318-4848 with a similar source, CI Cameleopardis.
\end{abstract}


\section{Introduction}
Since 2002, \textit{INTEGRAL} (INTErnational Gamma-Ray Astrophysics Laboratory) has been observing the sky looking for gamma-ray sources of various nature. On top of significantly increasing the number of known X-ray binaries, \textit{INTEGRAL} was able to discover a new type of highly obscured supergiant High Mass X-ray Binaries (sgHMXB) as reviewed in \cite{chaty_Optical/infrared_2013} and in \cite{walter_high-mass_2015}. These peculiar binaries host either a neutron star (NS) or a black hole (BH) in orbit around an early type supergiant star. Depending on the configuration of the binary, the compact object can be fed through the intense stellar wind of its giant companion, or by Roche Lobe overflow. The study of such extreme objects is crucial for understanding both the environment of supergiant stars and the products of binary interaction.

IGR J16318-4848 is the first source detected by \textit{INTEGRAL}, and is one of the most absorbed sgHMXB in the Galaxy known to this day. Discovered on January 29, 2003 \citep{courvoisier_igr_2003} with \textit{INTEGRAL/IBIS} in the 15--40\,keV band, the X-ray column density is so high ($N_{\text{H}} \simeq2\times10^{24}$\NH,\,\citel{matt_properties_2003}, \citel{walter_INTEGRAL_2003}) that its flux drops drastically below 5\,keV. It is known to be a Galactic persistent X-ray source with reccurent outbursts that last up to $\sim$20 days.

\cite{filliatre_optical/near-infrared_2004} use optical and nIR spectra to derive an absorption of A$_V$=17.4\,mag, which is far greater than the line of sight value of 11.4, while still a hundred times lower than in X-rays. This leads the authors to suggest a concentration of X-ray absorbing material local to the compact object, and the presence of a shell around the whole binary absorbing optical/nIR wavelengths. The nIR spectrum in \cite{filliatre_optical/near-infrared_2004} shows many prominent emission features in the same way CI Cam does, the first HMXB to be detected with an sgB[e] companion. P-Cygni profiles and forbidden [\ion{Fe}{2}] lines, also present in the nIR spectrum, are the evidence of a complex and rich environment, local to the binary.

Later, \cite{kaplan_long-wavelength_2006} use photometry to show evidence of mid-infrared excess in IGR J16318-4848. The authors find that a $\sim$1000\,K blackbody can be fit to the mid-IR excess in the spectral energy distribution (SED), and while they could not further characterize the exact nature and amount of the emitting material, they associate it to the presence of warm dust around the central star. \cite{moon_rich_2007} provide \textit{Spitzer} spectra from 5 to 40\,\micron\, that reveal a rich environment composed of an ionised stellar wind, a lower density region giving birth to forbidden lines, a photodissociated region and a two-component circumstellar dust (T\,$>$700\,K and T$\sim$180\,K). The actual geometry of this component was yet unknown, and if organized spherically around the central star would not contribute significantly to the absorption, meaning a much colder component (T\,$<$100\,K) located at the outermost regions of the binary could be responsible for the extreme extinction. \cite{rahoui_multi-wavelength_2008} perform photometry on IGR J16318-4848 with VLT/VISIR and reach a similar conclusion, i.e. that warm circumstellar dust is responsible for the MIR excess.

However, \cite{ibarra_XMM-Newton/INTEGRAL_2007} suggest that the column density is inhomogeneous and that the  circumstellar matter could very well be concentrated in the equatorial plane, seen almost edge-on, hence the very high $N_{\text{H}}$. The outflow might then be bimodal, with a fast polar wind and a slow, dense equatorial outflow.

\cite{chaty_broadband_2012} use VLT/VISIR mid-IR along with NTT/SofI and \textit{Spitzer} spectra to fit the SED of IGR J16318-4848. They report the presence of an irradiated rim around the star at T$_{rim}$\,=\,3500--5500\,K and a warm dust component at T$_{dust}$\,=\,767\,K in the outer regions of the binary using models of Herbig AeBe forming stars, which have circumstellar material analogous to IGR J16318-4848.

\cite{jain_orbital_2009} suggest a possible 80\,d period based on \textit{Swift}-BAT and \textit{INTEGRAL} data. Recently, \cite{iyer_orbital_2017} provide the results of a long-term observation campaign on IGR J16318-4848 with \textit{Swift/BAT}. They derive an orbital period of 80.09\p 0.01\,d. The folded lightcurves reveal two distinct peaks separated by low intensity phases. Several flares are detected, and preferentially happen in the same phase as the main peak, which may indicate that the compact object is crossing a denser and more inhomogeneous medium. There is also a correlation between the intensity of the flares and the time gap between them. This is reminiscent of disk-fed systems, where the material is regularly depleted then restructured.

Concerning the nature of the compact object, several pointed observations with \textit{XMM-Newton} (\citel{walter_INTEGRAL_2003}, \citel{iyer_orbital_2017}), \textit{Suzaku} \citep{barragan_suzaku_2009} and \textit{NuSTAR} \citep{iyer_orbital_2017} show no detectable pulsation in IGR J16318-4848. The absence of radio jets reported by \cite{filliatre_optical/near-infrared_2004} could indicate the presence of a neutron star which poles do not cross our line of sight, which is what we assume in the rest of this paper.

\begin{table*}
\caption{Log of our observations.\label{tab:log}}

\begin{tabular}{llcccccrr}
\hline\hline\\[-1.5ex]
Target  & Status & RA    & Dec   & Date &  \multicolumn{2}{c}{Airmass} & \multicolumn{2}{c}{Exposure} \\
        &        & J2000 & J2000 & (UTC) & Start & End & VIS & NIR \\
\hline\\[-1.5ex]
IGR J16318-4848 & SCIENCE & 16:31:48.41 & -48:49:03.54 & 2012-07-08T01:00:42 & 1.129 & 1.108 & 1200\,s & 200\,s \\
HD 145412       & CALIB   & 16:13:11.73 & -49:53:05.71 & 2012-07-08T00:26:00 & 1.156 & 1.155 & 5\,s    & 5\,s   \\
\hline\\[-1.5ex]
\end{tabular}

\end{table*}

\section{Observations and data reduction}\label{sec_obs}
The observations of IGR J16318-4848 and standard star HD145412 were performed in July 2012 at the European Southern Observatory (ESO, Chile) under program ID 089.D-0056 (see summary in \tabref{tab:log}). Spectra from 300 to 2480\,nm were acquired on the 8-meter Very Large Telescope Unit 2 Cassegrain (VLT, UT2) on three different arms (UVB, VIS and NIR) of the X-Shooter instrument. Because of the high intrinsic absorption of the source, the UVB spectrum only allows to set an upper limit to the flux. According to the phase diagram provided in Fig.\,3 in  \cite{iyer_orbital_2017}, our X-Shooter observations took place at phase 0.944\p 0.005, which corresponds to a low-intensity phase in X-rays.

\subsection{Optical (UVB, VIS) and near-infrared (NIR) data}\label{subsec_data}

Optical (UVB) echelle spectra were obtained through a 0.5\arcsec$\times$11\arcsec slit giving a spectral resolution of $R=9700$ (31\kmpersec) over a spectral range of 300--560\,nm with a dispersion of 0.02\,nm per pixel. Four exposures of 300\,s were taken, for a total integration time of 1200\,s.

Optical (VIS) echelle spectra were obtained through a 0.7\arcsec$\times$11\arcsec slit giving a spectral resolution of $R=11400$ (26\kmpersec) over a spectral range of 533--1020\,nm with a dispersion of 0.02\,nm per pixel. Four exposures of 300\,s were taken, for a total integration time of 1200\,s.

Near-infrared (NIR) echelle spectra were obtained through a 0.6\arcsec$\times$11\arcsec slit giving a spectral resolution of $R=8100$ (37\kmpersec) over a spectral range of 994--2580\,nm with a dispersion of 0.06\,nm per pixel. Twenty exposures of 10\,s were taken, for a total integration time of 200\,s.

All the acquisitions followed the standard ESO nodding pattern. The data reduction was performed with ESOReflex, using the dedicated X-Shooter pipeline. It consists of an automated echelle spectrum extraction along with standard bias, dark and sky subtraction along with airmass correction. Median stacking was used to add individual spectra in order to correct for cosmic rays.

The wavelength calibration was done during the reduction using calibration lamp frames and OH sky lines (NIR). The RMS of the solution is 0.003\,nm and 0.009\,nm for VIS and NIR spectra respectively.

The spectral response was obtained along with the flux calibration using the standard star spectrum (see \ref{subsec_resp}). Telluric absorption features were corrected using Molecfit \citep{kausch_molecfit_2015,smette_molecfit_2015}, a software that fits atmospheric features using a radiation transfer code and various parameters from the local weather.

\subsection{Detector response and flux calibration}\label{subsec_resp}
We use the spectrum of the standard star HD145412 to extract a response curve that allow us to both correct for spectral response and derive a flux calibration. The flux $F_\nu(\nu)$ in the standard star spectrum can be written as follows:
\begin{equation}
F_\nu(\nu) = R_\nu \times \left(F_{feat} + \left(\frac{R_*}{D_*}\right)^2B_\nu(\nu,T) \right) \times A_\nu \times A_\nu^{atm}
\end{equation}

where $R_\nu$ is the response curve we need to isolate. $F_{feat}$ is the flux of the emission and/or absorption features of the standard star, and $B_\nu(\nu,T)$ its blackbody continuum. R$^*$ and D$^*$ are the radius of the star and its distance, $A_\nu$ and $A_\nu^{atm}$ are respectively interstellar and atmospheric absorption.

Firstly, $A_\nu^{atm}$ is computed by Molecfit. In optical, absorption mainly comes from H$_2$O vapour and O$_2$, while CO$_2$, H$_2$O and CH$_4$ dominate in near-infrared.

Secondly, $A_\nu$ is computed using the formula in \cite{cardelli_relationship_1989} for optical and near-infrared. For HD\,145412 we use the following values found in \cite{morales_duran_Rv_2006}: A$_V$ = 0.77 and R$_V$ = 3.208.

Thirdly, the term $\left(\frac{R_*}{D_*}\right)^2B_\nu(\nu,T_{eff})$ is obtained by fitting a blackbody emission to the SED of HD145412, where:

\begin{equation}
B_\nu(\nu,T_{eff}) = \frac{2h\nu^3}{c^2}\frac{1}{exp(\frac{h\nu}{kT_{eff}}-1)}
\end{equation}

The \textit{Gaia} archive \citep{collaboration_gaia_2018} provides a parallax of 5.48$\pm$0.05\,mas for HD145412, which corresponds to a distance D$^*$=183$\pm$2\,pc. Using the calibrated photometric points available in the litterature (see \tabref{tab:phot_dat}), we find a blackbody of temperature T$_{eff}$ = 8615\p 133\,K and radius R$^*$ = 6.1\p 0.1\,R$_{\odot}$. Considering the uncertainties of the two parameters T$_{eff}$ and R$^*$, our flux calibration uncertainty varies from 8\% to 4\% on the full spectral range (533--2478\,nm).

\begin{table}[h]
\begin{center}
\small
\caption{Photometric data for HD 145412.\label{tab:phot_dat}}
\begin{tabular}{rccl}
\hline\hline\\[-1.5ex]
Wavelength & Flux & Flux err. & Reference \\
(nm)       & (Jy)         & (Jy)       &        \\
\hline\\[-1.5ex]
357.06 & 2.240 & 0.245 & UVB, Mermilliod 1991$^a$ \\
428.00 & 6.384 & 0.088 & Tycho$^b$ \\
437.81 & 6.261 & 0.685 & UVB, Mermilliod 1991 \\
534.00 & 6.889 & 0.063 & Tycho \\
546.61 & 6.483 & 0.709 & UVB, Mermilliod 1991 \\
585.76 & 5.519 & 0.604 & \textit{Gaia} DR1$^c$ \\
1\,235.00 & 4.595 & 0.097 & 2MASS$^d$ \\
1\,662.00 & 3.022 & 0.086 & 2MASS \\
2\,146.50 & 2.202 & 0.183 & DENIS$^e$ \\
2\,159.00 & 2.079 & 0.046 & 2MASS \\
3\,352.60 & 0.950 & 0.081 & WISE$^f$ \\
3\,507.51 & 0.880 & 0.046 & GLIMPSE$^g$ \\
4\,436.58 & 0.523 & 0.023 & GLIMPSE \\
4\,602.80 & 0.565 & 0.016 & WISE \\
5\,628.10 & 0.356 & 0.010 & GLIMPSE \\
7\,589.16 & 0.207 & 0.004 & GLIMPSE \\
8\,228.36 & 0.188 & 0.004 & AKARI$^h$ \\
11\,560.80 & 0.088 & 0.002 & WISE \\
22\,088.30 & 0.030 & 0.003 & WISE \\
\hline
\end{tabular}
\end{center}
\begin{small}
$^a$: \cite{mermilliod_vizier_2006}, $^b$: \cite{hog_tycho-2_2000}, $^c$: \cite{collaboration_gaia_2016}, $^d$: \cite{skrutskie_two_2006}, $^e$: \cite{epchtein_vizier_1999}, $^f$: \cite{wright_wide-field_2010}, $^g$: \cite{churchwell_spitzer/glimpse_2009}, $^h$: \cite{ishihara_akari/irc_2010}.
\end{small}

\end{table}

Finally, each individual spectral features F$_{feat}$ of the standard star are fitted with relation to the local continuum and subtracted. These features mostly come from hydrogen series in absorption (Balmer, Paschen and Brackett) and are lorentzian-shaped.

Because of the various residuals of feature fitting (both stellar and atmospheric) and poor signal-to-noise ratio towards the blue part of the specra, we apply a median filter with a 21\,pixel window width to the cleaned response curve. This allows us to obtain a smoother response at small scales while keeping its overall shape. Boundary effects of such filter on the edge of the detectors can be neglected behind signal-to-noise ratio drop at the edges and the coverage of both VIS and NIR arm in the region 994--1020\,nm. A normalized version of the final optical (VIS) to near-infrared (NIR) spectrum is shown in \figref{fig:fullspec}. As for UVB, the standard deviation of the data in the middle of the wavelength range (400--475\,nm) provides an upper limit on the flux of 4.7\tentothe{-6}\,Jy.

\section{Spectral feature analysis}

\begin{figure*}

\fig{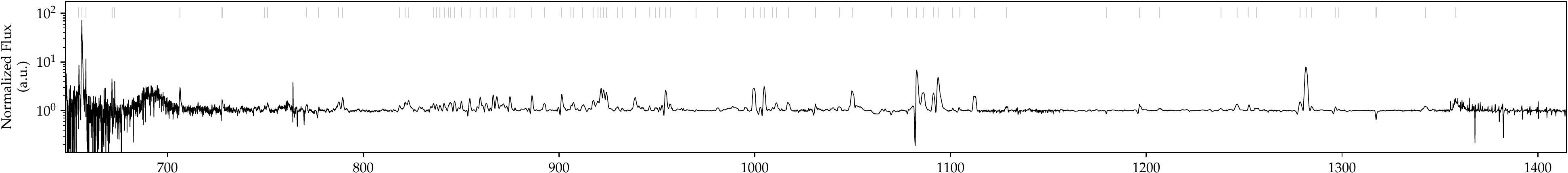}{\textwidth}{}

\fig{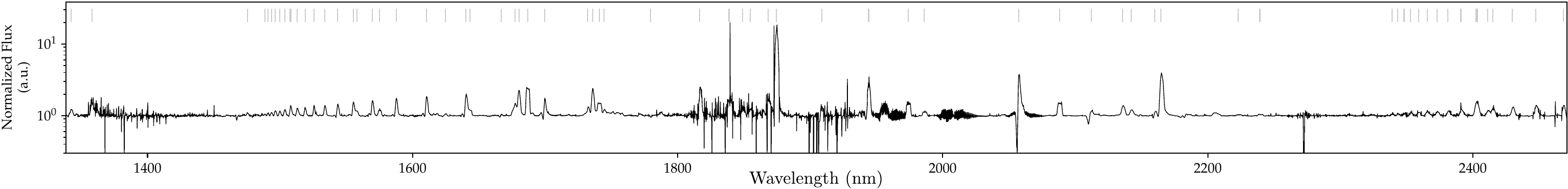}{\textwidth}{}

\caption{Optical to near-infrared X-Shooter spectrum of IGR J16318-4848. Identified lines are indicated with grey ticks.\label{fig:fullspec}}
\vspace{0.2cm}
\end{figure*}

Here we describe the features in the X-Shooter spectrum (see \ref{fig:fullspec}), and derive the associated parameters that we will discuss in Sect. \ref{sect:spec_interpretation}. The lines are identified using previous studies on IGR J16318-4848 \citep{filliatre_optical/near-infrared_2004} or studies on similar P-Cygni, sgB[e] or even T-Tauri sources (\citel{edwards_forbidden_1987}, \citel{hillier_optical_1998}, \citel{hynes_spectroscopic_2002}, \citel{clark_supergiant_2013}). We mostly detect hydrogen and helium in emission, along with iron and other metals. All the velocity shifts mentioned are given in the heliocentric restframe.

\subsection{Hydrogen lines}
The hydrogen lines are by far the most prominent features in the X-Shooter spectrum. The Paschen, Bracket and Pfund series are visible up to quantum numbers of $\sim$ 20--25, above which they become too faint and/or blended with one another. Balmer's \Halpha\,is also present, however its local background is under the detection limit; all values derived on this line will thus be lower limits. The list of hydrogen lines is available in \tabref{tab:line:hydrogen}.

The Pashen and Brackett lines show blueshifted absorption features, which is reminicent of P-Cygni profiles, although Pfund lines do not show such features. In the rest of the paper, the P-Cygni velocities are derived using the centers of each components (emission, absorption), and thus yield a lower limit on the expansion velocity of the medium.
Fitting the hydrogen P-Cygni profiles with a double-gaussian returns a mean velocity difference between emission and absorption of 264.8\p 4.3\kmpersec. The full-width at half-maximum (FWHM) of the emission component is 340.7\p 4.5\kmpersec\, in average, while it is of 204.9\p 9.4\kmpersec\,for the absorption component. The center of the emission line is blueshifted at -49\p 20\kmpersec\, in average. Several outliers may impact the estimate of the blueshift because of poor signal-to-noise ratio or a line profile that deviate from a single gaussian (discussed in \ref{subsect:Hmodel}); at this point, median statistics may give a more realistic velocity estimate of -48\p 10\kmpersec.

\begin{figure}[h]
\begin{minipage}[t]{.45\textwidth}
\fig{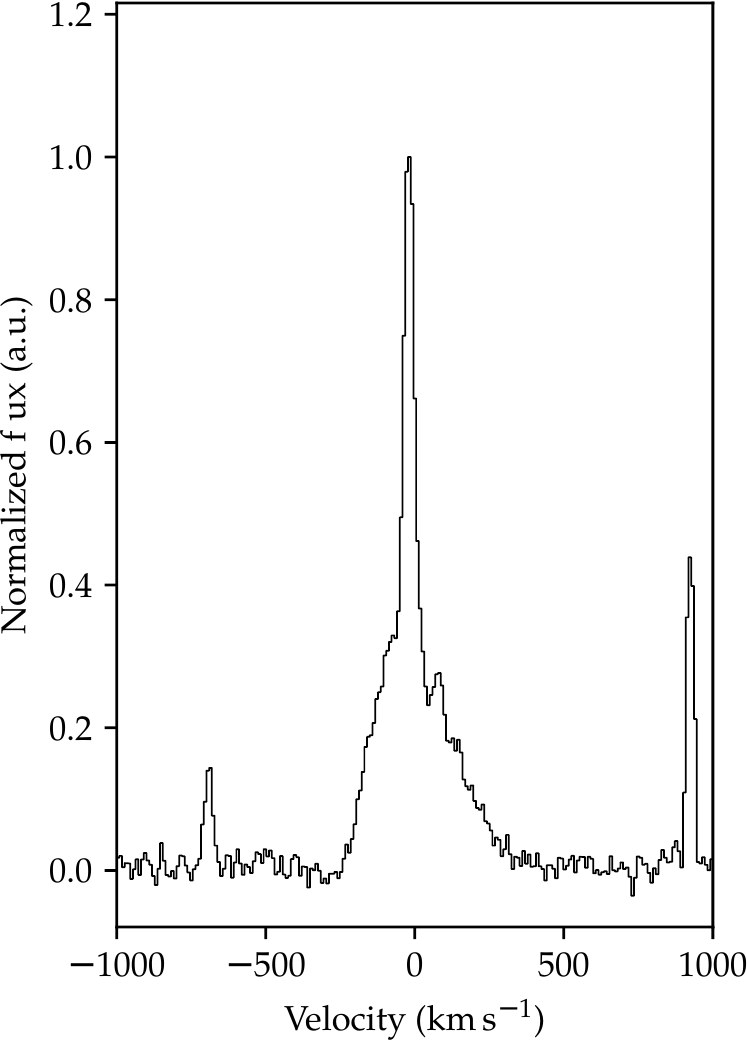}{.45\textwidth}{(a)}\label{fig:line:ha:before}
\fig{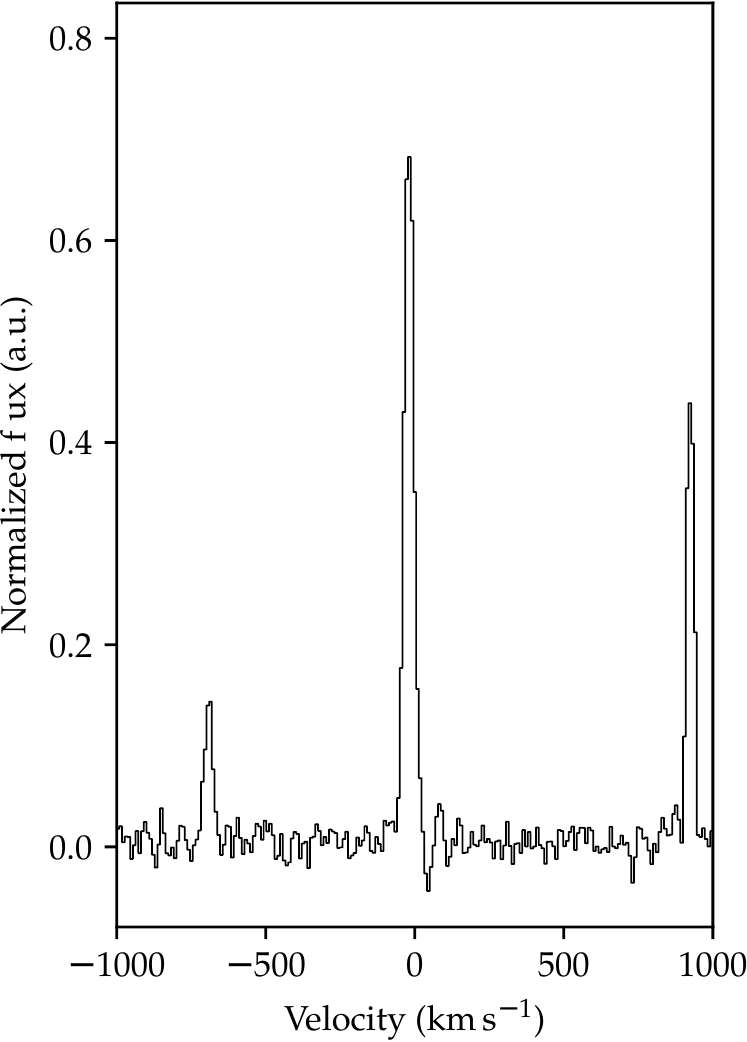}{.45\textwidth}{(b)}\label{fig:line:ha:after}
\caption{Evidence for the double component in \Halpha. The line is plotted (a): before and (b): after subtracting the average \ion{H}{1} P-Cygni profile at a velocity of 264.8\kmpersec.\label{fig:ha}}
\end{minipage}
\hfill\vspace{0.2cm}
\begin{minipage}[t]{.45\textwidth}

\fig{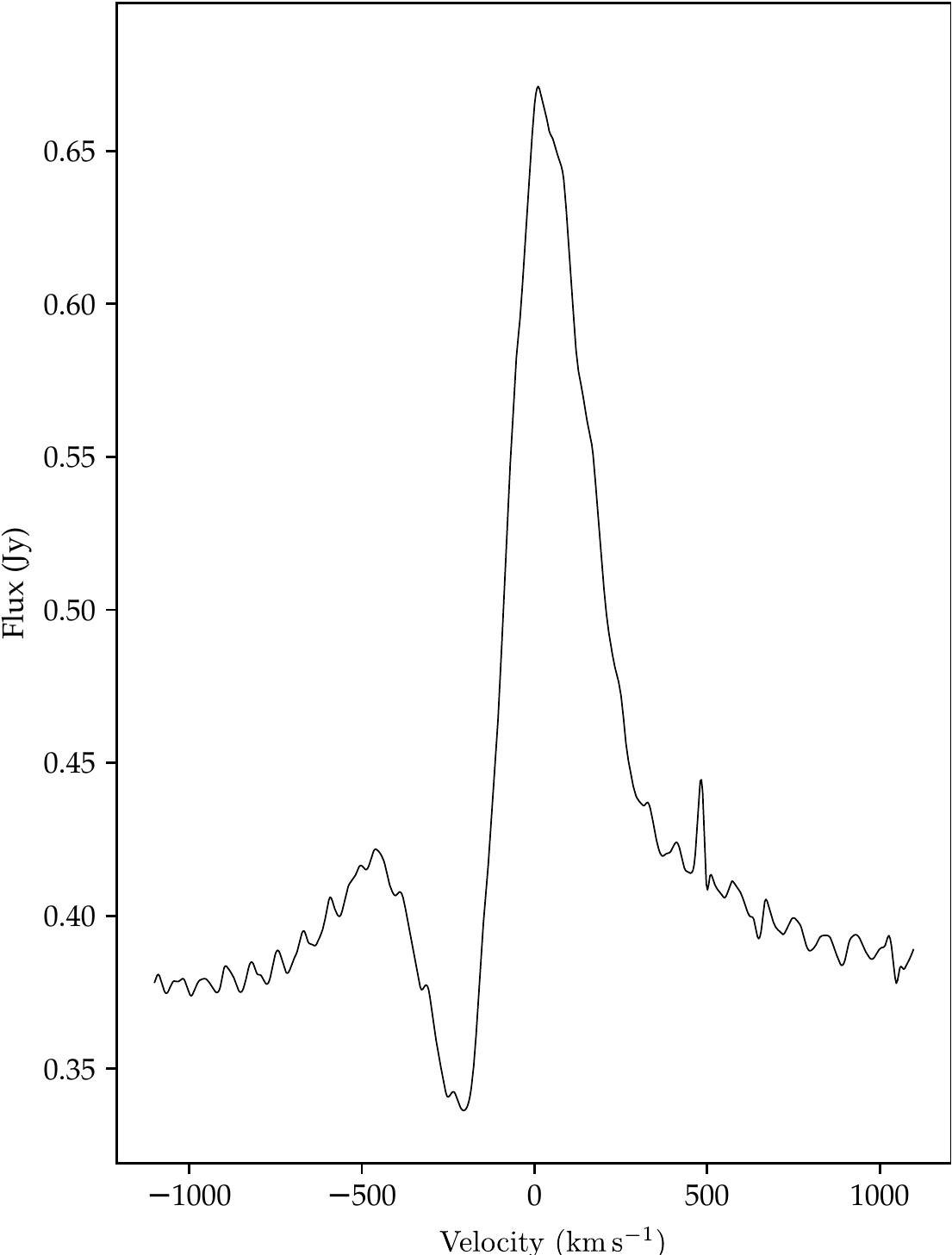}{.45\textwidth}{(a) \ion{He}{1} $\lambda$1700\,nm}
\fig{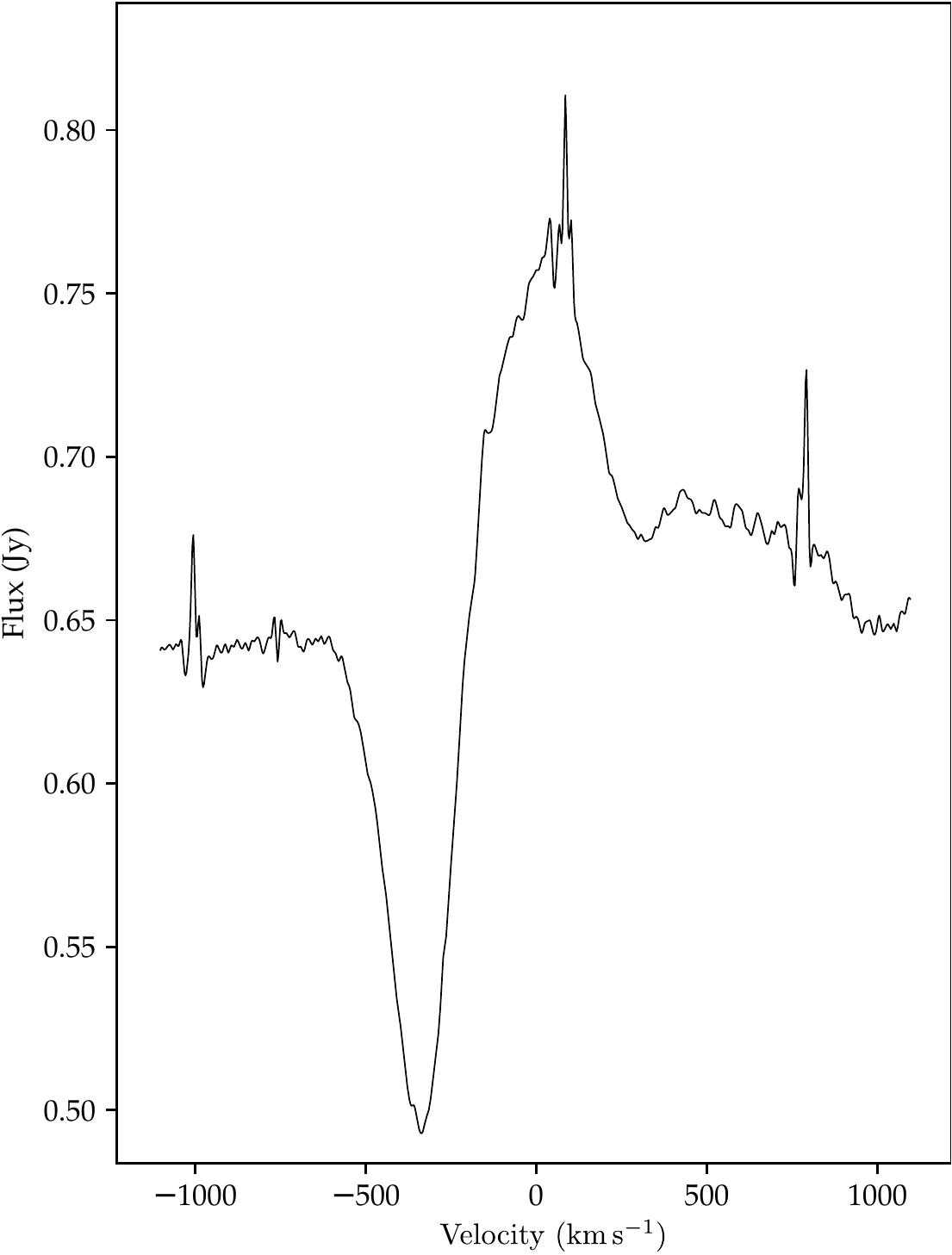}{.45\textwidth}{(b) \ion{He}{1} $\lambda$2112\,nm}

\caption{(a)\,: helium profile with extra broad emission wings. (b)\,: helium profile with strong P-Cygni absorption.}
\label{fig:line:hepcyg}
\end{minipage}
\end{figure}

Balmer \Halpha\,shows a thin emission line on top of a P-Cygni profile. This is highlighted by subtracting the average P-Cygni profile to the \Halpha\,line (see \figref{fig:ha}). The remaining line is then fitted with a single gaussian. We estimate its width at 25.6\p 3.3\kmpersec, an order of magnitude lower than the other hydrogen lines in the spectrum (see Tab.\,\ref{tab:line:hydrogen}).

\subsection{Helium lines\label{subsect:helines}}
All helium lines are detected in emission and are atomic \ion{He}{1} (see \tabref{tab:line:helium}). They show blueshifted absorption features like hydrogen lines. Some \ion{He}{1} lines may have extra components on top of the common P-Cygni profile. For instance, \ion{He}{1} $\lambda$1700\,nm  and $\lambda$2058\,nm (see \figref{fig:line:hepcyg}) appear to have an extra wide emission component. The P-Cygni emission and absorption components in these transitions are separated by 150\p 30\kmpersec\, in average. The mean FWHM of the emission is 290\p 20\kmpersec, while it is 250\p 10\kmpersec\, for the absorption. The third, wide component can be reproduced by an emission line of width 840\p 20\kmpersec. Its center is compatible with the center of the P-Cygni emission line in both $\lambda$1700\,nm and $\lambda$2058\,nm transitions.

The $\lambda$2112\,nm line has a rather particular profile (\figref{fig:line:hepcyg}b). Given that the median ratio of equivalent width between emission and absorption is 1.6 for helium lines, it is of 1.2 for this particular line, making the P-Cygni absorption stronger in comparison. We note that the emission line is centered at a similar velocity as other helium lines (-119\kmpersec). However, we measure a greater P-Cygni velocity at -352\kmpersec.

\subsection{Metallic lines}
\subsubsection{Iron flat-topped lines}
\begin{figure}[h]
\fig{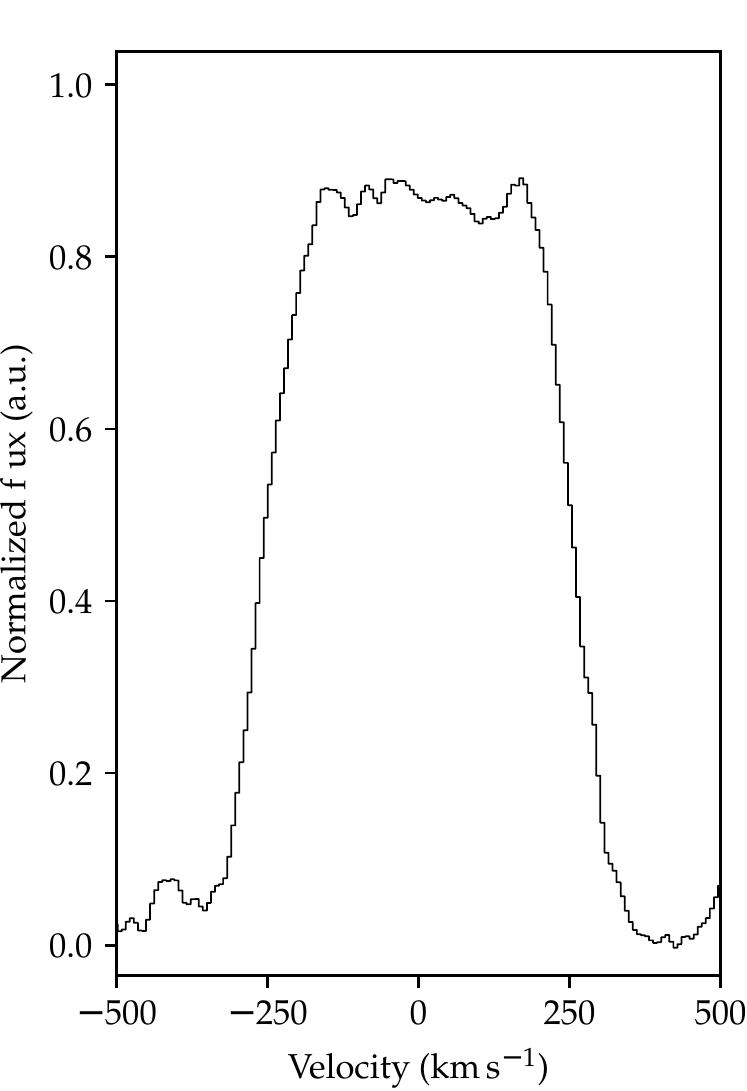}{.3\textwidth}{}
\caption{Average flat-top profile of \ion{Fe}{2} lines in IGR J16318-4848.\label{fig:flattop}}
\end{figure}

Lines associated to \ion{Fe}{2} (and [\ion{Fe}{2}], although much fainter) present a unique, flat-topped profile with narrow symetrical wings. We show the average \ion{Fe}{2} profile in Fig.\,\ref{fig:flattop}. To fit these lines, we use a model which consists of a rectangle function convolved with a gaussian. All the identified flat-topped lines from \ion{Fe}{2} and [\ion{Fe}{2}] are listed in Tab.\,\ref{tab:line:iron}.

For the allowed transitions of \ion{Fe}{2}, the average half-width of the rectangle component is 250\p 20\kmpersec, and 81\p 22\kmpersec\, for the half-width of the gaussian broadening. The average heliocentric velocity is -75\p 7\kmpersec.

For the forbidden transitions [\ion{Fe}{2}], the average half-width of the rectangle component is 285\p 6\kmpersec, and 43\p 6\kmpersec\, for the half-width of the gaussian broadening. The average heliocentric velocity is -47\p 6\kmpersec.

\subsubsection{Other metals}

Forbidden [\ion{N}{2}], [\ion{O}{1}] and [\ion{S}{2}] can be found close to \Halpha. They all have very small widths ($<$25\kmpersec), and their average blueshift is -32\p 1\kmpersec. These two parameters are significantly different from all the others lines visible in the spectrum.

\ion{Mg}{2} doublets are found across the optical to near-infrared spectrum, with intensities comparable to helium or even hydrogen lines. They are compiled in \tabref{tab:line:mg}. Without taking into account the polluted lines with bad signal, their average FWHM is 339\p 40\kmpersec, and average blueshift of $-100$\p 30\kmpersec.

\subsection{Diffuse interstellar bands}\label{subsect:dibs}
\begin{figure}[h!]

\fig{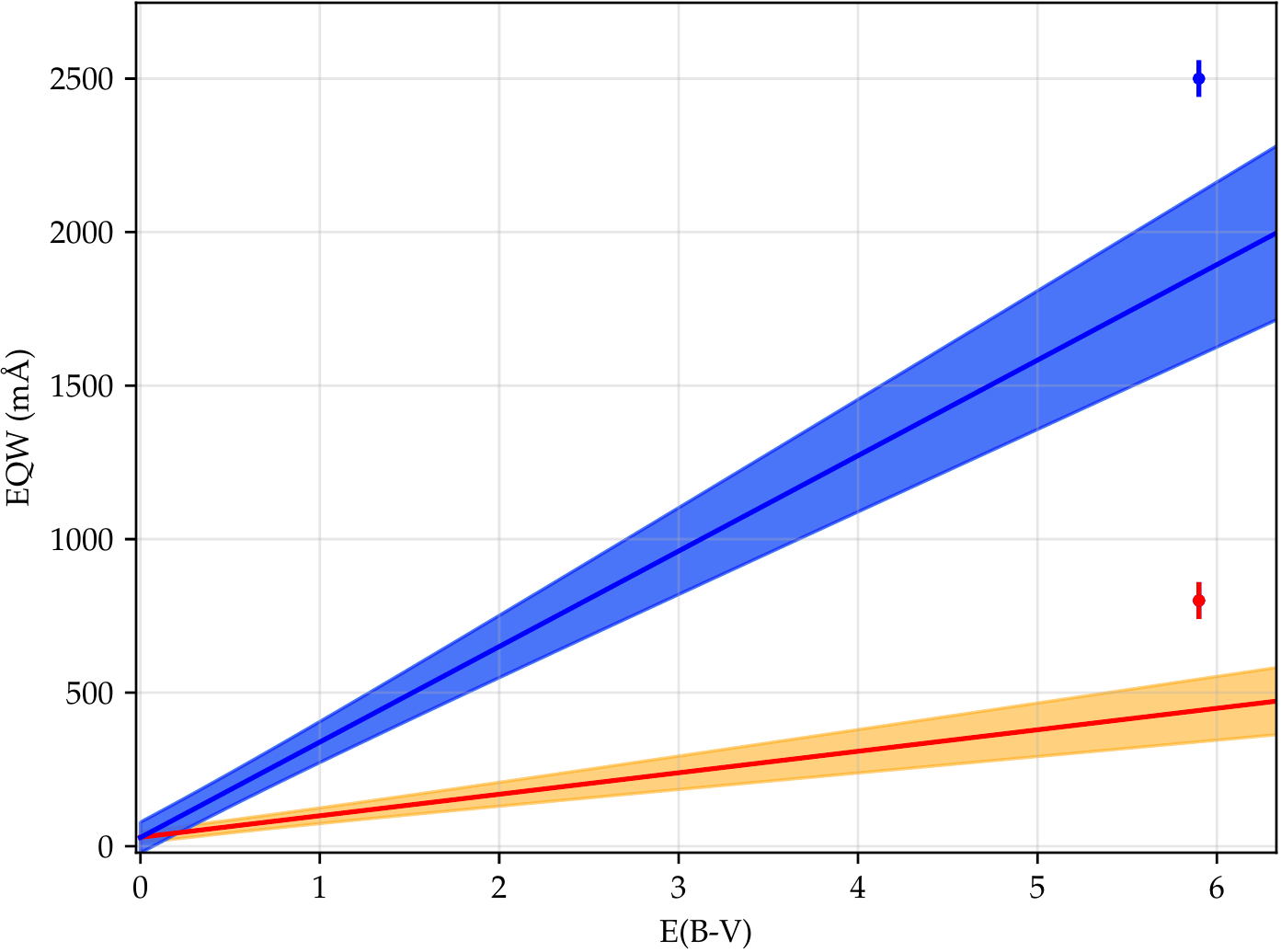}{.45\textwidth}{}
\caption{Correlation of the equivalent widths of DIBs with extinction (lines, adapted from \cite{cox_VLT/X-Shooter_2014}) versus our measurements (dots). Red: $\lambda$1180\,nm, blue: $\lambda$1317\,nm.\label{fig:dibs}}

\end{figure}

A set of Diffuse Interstellar Bands (DIBs) in X-Shooter spectra are presented in \cite{cox_VLT/X-Shooter_2014}. The authors show a correlation between the equivalent widths of the DIBs with the absorption in the line of sight. We measured the equivalent width of the DIBs we detected in the spectrum of IGR J16318-4848 (see \tabref{tab:dibs}). Our EQW measurements of DIBs are all compatible with A$_V>$\,10.9 according to the correlation coefficients in \cite{cox_VLT/X-Shooter_2014} (see comparison in \figref{fig:dibs}). If we compute the absorption with the relation they derived, we obtain A$_V$ = 38.6$\pm$2.2 for $\lambda$1180\,nm and A$_V$ = 23.6$\pm$5.6 for $\lambda$1317\,nm. The first is not a realistic estimation, as the correlation domain does not go above A$_V$ = 11 (E$_{B-V}\sim$3.5) and we suggest it is no longer valid at higher absorption values. On the other hand, the value computed for the $\lambda$1180\,nm transition is more realistic, although the correlation has a single point above A$_V$ = 11 and as such the confidence interval is rather large. We note that the value is still compatible with the one derived \cite{chaty_broadband_2012} of 18.3\p 0.4. However, considering the high uncertainty of the correlation in this domain, we suggest that we cannot improve the estimation of the absorption, and that it is preferable to keep the value of A$_V$=18.3.

\begin{table}[h]
\caption{List of diffuse interstellar bands found in the spectrum of IGR J16318-4848 compared to DIBs found in \cite{cox_VLT/X-Shooter_2014}\label{tab:dibs}}
\begin{tabular}{cccc}
\hline\hline\\[-1.5ex]
$\lambda_{DIB}$ (nm) & $\lambda_{fit}$ (nm) & FWHM (nm) & EQW (nm)\\
\hline\\[-1.5ex]
1069.7 & 1069.687 & 0.738 & 0.13 \\
1078.0 & 1077.899 & 0.375 & 0.07 \\
1179.7 & 1179.560 & 0.500 & 0.09 \\
1317.5 & 1317.385 & 0.664 & 0.24 \\
1780.3 & 1779.843 & 1.120 & 0.07 \\
\hline
\end{tabular}

\end{table}

\section{Spectral feature interpretation}\label{sect:spec_interpretation}
In this section we use all the information derived from the spectrum and confront it to the previously inferred geometry of IGR J16318-4848 to discuss the origin of the various line profiles.

\subsection{Hydrogen to helium abundance ratio} \label{subsect:HHe}
Following the method presented in \cite{allen_near-infrared_1985} to derive the hydrogen to helium abundance ratio, we use the intensity ratio of \ion{He}{1} $\lambda$1700\,nm and $\lambda$1200\,nm versus \brgamma. This method provides a lower limit on the abundance, since it is probing H$^+$ that recombines into \ion{H}{1} (same for helium). With an absorption of A$_V$=18.3 magnitudes, the first ratio \ion{He}{1} $\lambda$1700\,nm/\brgamma\, gives 0.248 while the \ion{He}{1} $\lambda$1200\,nm/\brgamma\, ratio provides 0.296. Both measurements are compatible with an evolved environment, as it is at least 3.7 times the solar abundance.

This could very well come from the properties of the birthplace of IGR J16318-4848. Especially, the disc could be at least partly formed from leftover material after the formation of the system. However, the first supernova event could also have fed the medium around the binary with heavy elements. Also, the central supergiant star likely emits a strong wind, that could also participate in enhancing the environment with helium and metals, although the modeling of the stellar atmosphere and wind we present later in Sect.\,\ref{sect:power} may suggest that the helium mainly comes from the star itself.

\subsection{Further investigating the profile of hydrogen lines\label{subsect:Hmodel}}
Hydrogen lines with sufficient SNR are visibly different from a regular single gaussian profile. This is particulary the case for \brgamma\, and \ion{H}{1}\,$\lambda$1280\,nm, which have symetric distortions around both sides of the emission line.

In the study of the supergiant A[e] binary HD\,62623, \cite{millour_imaging_2011} present spectro-interferometric data that is compatible with the \brgamma\, line originating from the inner rim of the equatorial dusty disk of the system, which produces a double-peaked profile that probes the orbital velocity of the medium.

\begin{figure}
\fig{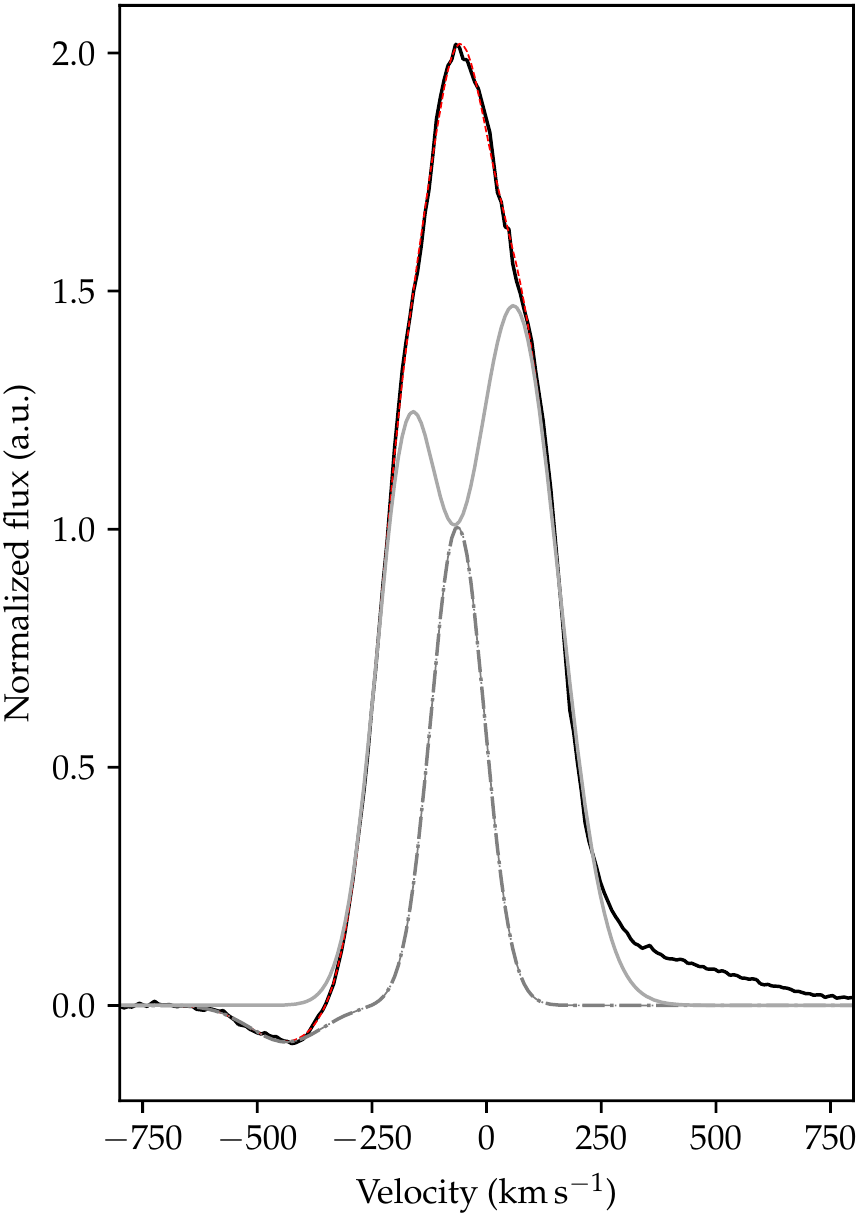}{.22\textwidth}{\brgamma\, $\lambda$2160\,nm}\label{fig:line:brgamma_model}
\fig{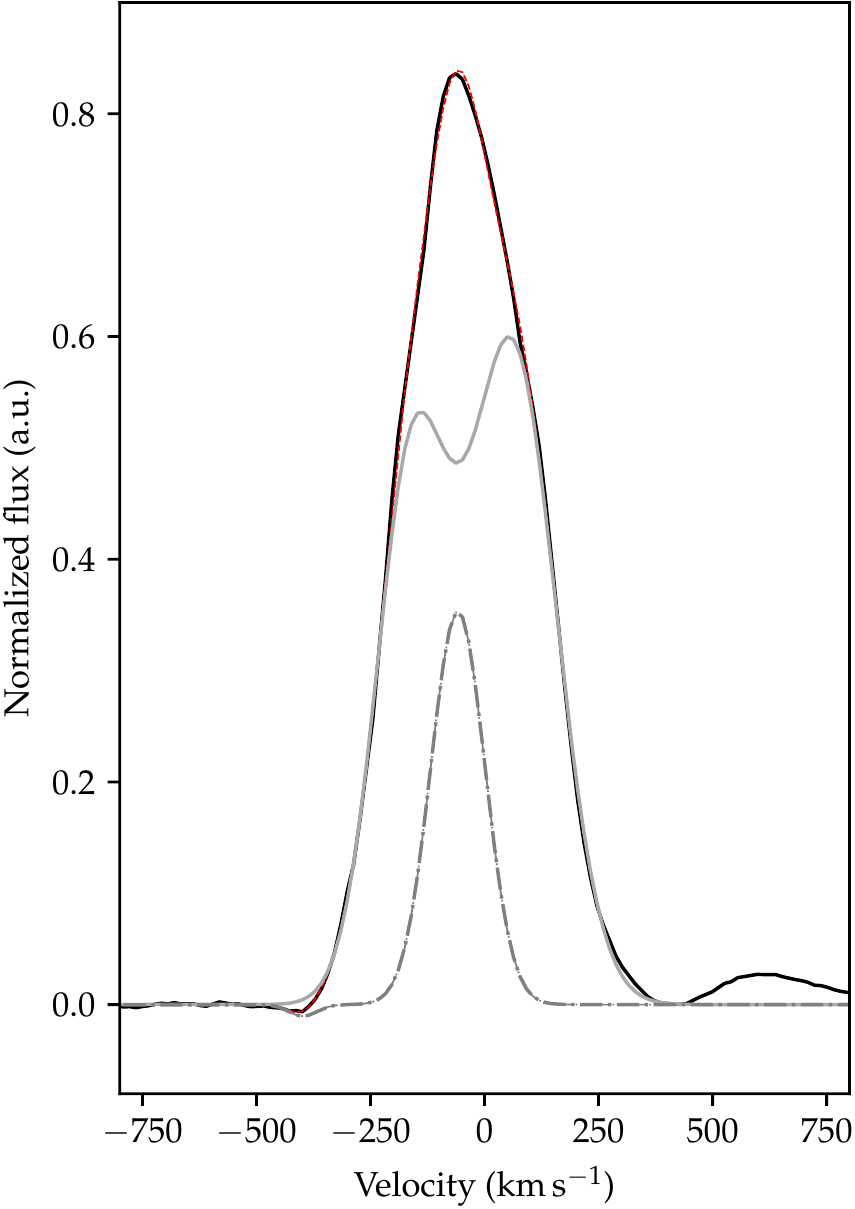}{.22\textwidth}{\ion{H}{1} $\lambda$1280\,nm}\label{fig:line:HI_model}
\caption{Hydrogen line profiles fitted with a P-Cygni profile on top of a double peaked emission from a keplerian rim. The continuum-subtracted data is in black, the model in red, and the individual components in dotted grey.\label{fig:line:Hmodel}}

\end{figure}

We thus fitted a double-peaked emission to the P-Cygni profile of \brgamma\, and \ion{H}{1}\,$\lambda$1280\,nm (\figref{fig:line:Hmodel}). It is possible to reproduce the data with a double peak, with the condition of adding an extra emission component of lower amplitude centered on the double-peaked profile.

For \brgamma\, (\figref{fig:line:brgamma_model}), the P-Cygni velocity is measured to be 373$\pm$11\kmpersec. The double peak is centered at -69\p 5\kmpersec\, and separated by 232\p 10\kmpersec, which corresponds to an orbital velocity vsin(i)=116$\pm$5\kmpersec. The extra emission component is centered at -76\p 5\kmpersec. Residuals on the red part of the line may correspond to the wing of the \brgamma\, line due to electron scattering.

For \ion{H}{1} $\lambda$1280\,nm\, (\figref{fig:line:HI_model}), the P-Cygni velocity is measured to be 340$\pm$12\kmpersec. The double peak is centered at -63\p 7\kmpersec\, and separated by 219\p 13\kmpersec, which corresponds to an orbital velocity vsin(i)=110$\pm$7\kmpersec. The extra emission component is centered at -71\p 5\kmpersec. Residual on the red part of the line is from an extra helium emision line also presenting a P-Cygni profile.

While the two P-Cygni velocities we derive are not fully compatible, the double peak structures are similar in both transitions and provide a mean orbital velocity of vsin(i)=113\p 4\kmpersec. The center of the double peak structure is located at -66\p 4\kmpersec. We note that for both lines, the center of the double-peaked profile is consistent with the center of the extra emission component.

The double-peaked profile and the P-Cygni absorption are features expected in the geometry of IGR J16318-4848, however we have yet to explain the origin of the extra emission component needed to reproduce the data. Because its center is compatible with the one of the double-peaked profile, we would first suggest that it comes from circumbinary material. But it could very well be a coincidence, and in fact originate from the central star, as the maximum orbital velocity amplitude of the star is 6--30\kmpersec\, (for a central star of 25--50\Msun\, and a compact object of 1.4--10\Msun\, in circular keplerian orbit seen purely edge-on).

Concerning the asymmetry of the double-peaked component, we can quantify it by calculating the peak-to-peak amplitude asymmetry defined as A =$|\Delta I|$/$\bar{1}$, where $I$ is the amplitude of the peaks in which continuum was subtracted. This results in an amplitude asymmetry of 24\p 7\% for \brgamma\, and 21\p 9\% for \ion{H}{1}\,$\lambda$1280\,nm.

This asymmetry can either come from a deviation from a circular orbit, or from the interaction between the compact object with the rim, bringing its heated material closer into the central cavity in the form of a wake. One way to pinpoint the origin of the asymmetry would be to observe the line profile change overtime. If the rim is affected by the compact object, then the signal should be modulated by the period of the binary ($\sim$80\,d, \cite{iyer_orbital_2017}). If ellipticity is responsible, then the modulation should come from the precession of the circumbinary material, which would be at a longer, superorbital period (see e.g. \citel{charles_optical_2003}). Although our current spectroscopic data does not allow us to chose one case over another, the results on the broadband spectral energy distribution discussed in Sect.\,\ref{sect:SED} allow cases in which the compact object orbits close enough to the rim so that its gravitational pull is of the same order of magnitude as the central star. This could set up the conditions for the rim to be modulated by the periodic passing of the compact object.

\subsection{Origin of the narrow emission lines}
\begin{figure}[h]
\fig{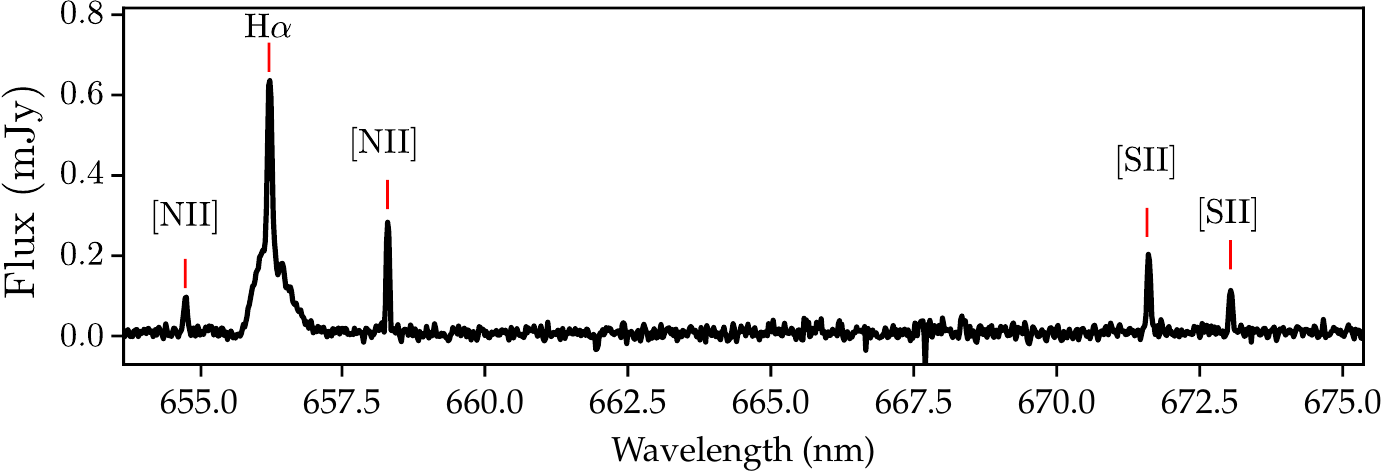}{.45\textwidth}{}
\caption{X-Shooter spectrum around the region containing narrow lines, indicated by red ticks.}
\label{fig:forbidden_closeup}
\end{figure}

As shown by the previous results on SED fitting \citep{chaty_broadband_2012}, the nIR continuum is dominated by the emission of the irradiated rim. In our X-Shooter spectrum, the nIR features are typically broadened at velocities in the order of FWHM=250\kmpersec, which could be associated to the orbital velocities of the medium around the center of mass. This is the case for all the different elements we identify (H, He, Mg, Fe...), and is further supported by the discussion in Sect.\,\ref{subsect:Hmodel}. However, below 700\,nm we find features that are noticeably different, mostly because of their much narrower width, as shown in Fig.\,\ref{fig:forbidden_closeup}.

\subsubsection{\Halpha\, line}

The most prominent of these features is the second component of the \Halpha\, line, which lies on top of a P-Cygni line profile analogous to other \ion{H}{1} lines. Its center is shifted at -33\kmpersec, and its FWHM is 25.6\kmpersec. While the width is definitely much smaller than all the other lines in the spectrum (by a factor $\sim$15), we cannot firmly tell if the velocity shift of the line center is significantly different from other hydrogen lines.

In the sgB[e] star RMS 82, \cite{seriacopi_envelopes_2017} report a slight depolarisation along the \Halpha\, line compared to its local continuum. This is compatible with the scenario where \Halpha\, is produced in a large volume around the star, thus suffering less scattering from the star's envelope, hence the lower polarisation.

This rules out the hypothesis that the narrow \Halpha\, line originates from the central star itself. It cannot originate from orbiting material close to the central star (i.e. the rim), because of the keplerian velocity being much larger than its width. Further away in the disk, where the width of the line could be compatible with the lower keplerian velocity, the medium is too cool and the conditions to form \Halpha\, are not met. If it were directly produced in the equatorial wind itself, we would expect it to be broadened at HWHM$\sim$v$_{p-cyg}$, close to 400\kmpersec. We can thus exclude an equatorial origin for this component.

This leaves only the polar wind of the central star, in analogy with the results in \cite{seriacopi_envelopes_2017}. If this is the case, it would suggest that the polar wind is collimated and seen almost edge-on (see discussion in Sect.\,\ref{subsect:inclination}).

\subsubsection{Forbidden metallic lines}
Five other narrow lines are found within the range 630--680\,nm. We identify them as the forbidden transitions from [\ion{O}{1}], [\ion{N}{2}] and [\ion{S}{2}]. We provide their characteristics in \tabref{tab:line:forbidden}, however because the continuum is under the detection limit in this part of the spectrum, these values should be taken with great caution. Their FWHM are all measured to be less than 25\kmpersec, and they are centered at -31\p 5\kmpersec\, in average.

We note that the same [\ion{O}{1}] and [\ion{N}{2}] transitions were identified in another sgB[e] star LHA 115-S 18 \citep{clark_supergiant_2013}, while the [\ion{S}{2}] lines were not. The significantly narrower profile of these lines indicate that they do not come from the same region as the other, wider lines that come from the rim.

A study from \cite{edwards_forbidden_1987} report the detection of all the aforementioned forbidden lines in what the authors argue to be the bipolar wind of T-Tauri stars. The authors also refer to a book by \cite{pottasch_planetary_1984} in which a relation between the ratio of [\ion{S}{2}] and the electron density of the medium is provided. Since the local continuum in our spectrum is not well defined, it is hard to have a good estimate of their equivalent width in IGR J16318-4848. With our data, the ratio I(673)/I(671) is 0.6\p 0.4, corresponding to an upper limit in the electron density of n$_e$\,$<$\,6$\times$10$^2$\,cm$^{-3}$.

However the presence of similar forbidden lines may be spurious. This is further supported by a recent spectroscopic study on a sample of sgB[e] stars \citep{maravelias_resolving_2018}. The authors report on the profile of forbidden emission lines, one of them being in common with IGR J16318-4848 ([\ion{O}{1}]\,$\lambda$630.0\,nm). The analysis reveals double-peaked profiles that indicate they originate from ring-like structures in orbit around the central star; for confirmed binaries, the profiles change over time. We do not resolve the profiles of the forbidden lines in the X-Shooter spectrum of IGR J16318-4848, so we cannot confirm if they indeed show similar structures. But their low velocity shift and width compared to the other lines is compatible with them coming from the equatorial plane of the dusty disk, as long as they originate more than 120--260\,au away from the central star, which is the closest keplerian orbit they could have considering their width.

\subsubsection{Nebular origin of narrow lines}

Another possibility is for all the optical narrow lines to come from foreground emission of the interstellar medium. \cite{ferland_pumping_2012} measure [\ion{O}{1}] line widths in the Orion Nebula to be 12.6\p 2.4\kmpersec\, in FWHM, which is the same order of magnitude as our FWHM measurement of the $\lambda$630\,nm [\ion{O}{1}] line in the IGR J16318-4848 optical spectrum. If the narrow lines are nebular, this would set the minimal distance of IGR J16318-4848 to be 2.4\p 0.3\,kpc (see the discussion on distance in Sect.\,\ref{sect:distance}), as the velocity shift of those lines is compatible with the velocity of a star forming region located close to the line of sight, in the Carina-Sagitarius arm.

Given the widths of the optical narrow lines are all very close to the instrumental resolution of X-Shooter, we suggest that performing high-resolution spectroscopy (R$>$20\,000) on the forbidden lines and the \Halpha\, line is necessary to clear up their origin.

\subsection{Origin of the flat-topped lines}
In \cite{bertout_line_1987}, different ways of forming flat-topped lines are discussed, and all concern material in spherical expansion. An optically thin medium for which the turbulent velocity is much smaller than the macroscopic outflow velocity (i.e. the expansion velocity) can generate flat-topped lines. In the case of an optically thick medium, the turbulent velocity has to be much greater than the outflow velocity to form the flat top. In the first case, the half-width of the profile probes the expansion velocity; in the second case it probes the turbulent velocity.

Out of the fifteen iron lines we detect in the spectrum of IGR J16318-4848, twelve are permitted transitions while three are forbidden. The forbidden lines are much fainter than the former; because of that, it is possible that the  difference in their characteristics (blueshift, width...) can be attributed to noise rather than an actual intrinsic difference.

Because the forbidden [\ion{Fe}{2}] lines are very likely to be optically thin, regular \ion{Fe}{2} lines sharing the same properties means the medium itself is thin. This favours the case in which the expanding medium which gives rise to flat-topped lines in IGR J16318-4848 is optically thin, and we can estimate its terminal velocity to be 250\p 20\kmpersec. If we associate the broadening of the wings to the orbital motion of the medium emitting the lines, their keplerian distance to a 25--50\Msun\, central star would be 3.4--6.8\,au for a circular orbit.

\subsection{Inclination of the system}\label{subsect:inclination}
Several studies on IGR J16318-4848 provide hints towards a very high inclination system, however there is yet to have a quantitative value. According to \cite{matt_properties_2003}, the shape of the Fe and Ni K$\alpha$ lines can be best reproduced considering an absorbing medium with a small covering factor, i.e. a disk seen at high inclination.
Similarly, \cite{barragan_suzaku_2009} argue that despite the high absorption, the lines in the \textit{Suzaku} spectrum do not need Compton scattering to be reproduced, which is consistent with a non-spherical and inhomogeneous distribution of the absorbing material. In \cite{chaty_broadband_2012}, the authors report on additional absorption components from silicates, which could be either due to an inaccurate absorption law or auto-absorption from a circumstellar disk seen edge-on, at high inclination.

In our X-Shooter spectrum, another argument in favor of a high inclination angle comes from the \Halpha\, line. Its narrow component is likely to come from the polar wind of the central star. The profile shows no sign of deviation from a single gaussian, and its FWHM of 25\kmpersec\, favors a narrow opening angle. Because of the typically high velocity of such winds, a small deviation from an edge-on line of sight should greatly impact the shape of the profile. For a fiducial terminal velocity of 1000\kmpersec\, (order of magnitude in CI\,Cam polar wind), and considering our resolution of 26\kmpersec\, in optical, we would be able to detect the blue and red components of a double-peaked \Halpha, corresponding to the symetrical polar wind, if the inclination angle was $\sim$1.5\degr\, off the edge-on configuration (i.e. i$<$88.5\degr).

\section{Distance and X-ray luminosity}\label{sect:distance}
\cite{filliatre_optical/near-infrared_2004} estimated the distance to be between 0.9--6.3\,kpc using the typical bolometric luminosity and temperature of sgB[e] stars. Later, \cite{chaty_broadband_2012} used mid-infrared data with VLT/VISIR along with a Herbig Ae/Be model consisting of a hot star ($\sim$20\,000\,K), an irradiated rim ($\sim$5\,500\,K) and a dusty disk ($\sim$900\,K) compatible with a distance of $\sim$1.6\,kpc as derived in \cite{rahoui_multi-wavelength_2008}.

The spectral features in the X-Shooter spectrum hint that the local environment of the binary could be a strong contributor to the overall SED. If this is the case, the mIR data would thus probe the region where the irradiated rim contributes roughly as much as the outer dusty disk. The sheer size of these emitting regions could make them much brighter than the central star in mIR. This means the aforementioned distance of 1.6\,kpc is likely to be underestimated.

\subsection{Nearby Star Forming Regions}\label{subsect:SFR}

IGR J16318-4848 is an sgB[e] HMXB, and is thus a young object. It is likely to be located close to a SFR (Star Forming Region), in one of the spiral arms in the line of sight (see e.g. \citel{coleiro_distribution_2013}), as suggested in \cite{filliatre_optical/near-infrared_2004}.

\begin{figure}[h]
\fig{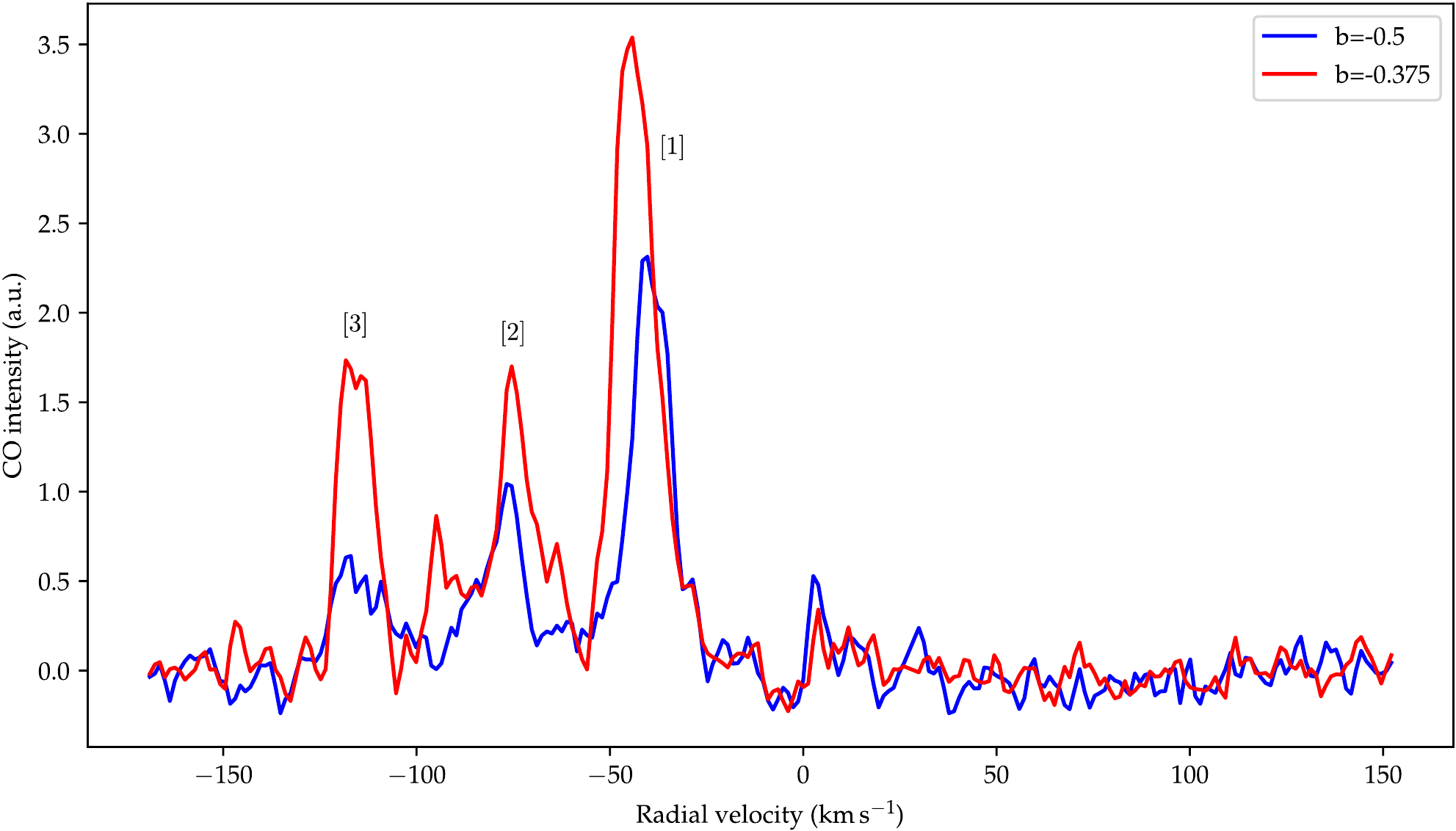}{.45\textwidth}{}
\caption{Radial velocities of CO regions from \cite{dame_Milky_2001} around the position of IGR J16318-4848.\label{fig:co_velocity}}
\end{figure}

We use velocity maps from \cite{dame_Milky_2001} around the position of IGR J16318-4848 (l=335.61, b=-0.447) to look for the radial velocity of the spiral arms in the line of sight. Two cuts around the position (b1=-0.5 and b2=-0.375) are available (see Fig.\ref{fig:co_velocity}). The two cuts show the same three regions with different velocities. We fit each region with a gaussian to derive their central velocity and standard deviation. We found that [1] peaks at -39\p 4.7\kmpersec, [2] peaks at -78\p 4.7\kmpersec, and [3] at -115\p 6.4\kmpersec.

We then retrieve the online data of \cite{russeil_Star-forming_2003} to find the distances of the star forming regions (SFR) we could associate to these regions. Two SFRs are found close to the line of sight. The first lies in the Carina-Sagitarius arm (l=334.7, b=-0.1) at 2.4\p 0.3\,kpc, has a radial velocity of -33\kmpersec\, and is likely to be associated with region [1] from Fig.\,\ref{fig:co_velocity}. The second is in the Scutum-Crux arm (l=335.9, b=0.2) at 4.9\p 0.2\,kpc, has a radial velocity of -78\kmpersec\, and is associated to region [2].

In terms of angular separation from the SFRs, IGR J16318-4848 is closer to the second (0.7\degr) than the first (1\degr); however considering the precision of the SFR coordinates, we reckon the difference in separation cannot be a decisive argument to associate IGR J16318-4848 to any of the two SFRs.

\subsection{Finding the best estimator of the radial velocity\label{subsect:radvel}}

Considering the masses of the central star and of the compact object are likely to be between 25-50\Msun\, and 1.4-10\Msun\, respectively, the maximum orbital velocity of the star around the center of mass is 29\kmpersec\, for a circular orbit. If we assume the narrow component of \Halpha\, comes from the polar wind of the star, then its shift of -33\kmpersec\, can yield a systemic velocity between -62 and -4\kmpersec. Without any prior knowledge of the full radial velocity curve, all the values of systemic velocity are equally probable whithin this range. We reckon this estimator is rather uncertain, and thus compatible with IGR J16318-4848 being associated to both SFRs [1] and [2].

If we instead consider the velocity shift from the center of the double peaked profile of hydrogen coming from the rim, we can assume that is is not affected by any significant orbital motion around the center of mass other than its keplerian orbit. Thus, it would provide a systemic velocity of -66\p 4\kmpersec, which favours an association with SFR [2] at 4.9\,kpc.

\subsection{Using the latest results from \textit{Gaia} DR2}

IGR J16318-4848 was observed by \textit{Gaia} and appears in the second data release. However, while its proper motion has been successfuly measured, the parallax provided in \textit{Gaia} DR2 is negative (p = -0.5\p 0.3\,mas). This value cannot be explained by the suggested zero-point systematic of -0.03\,mas, and thus cannot be used to derive a distance by simply inverting the parallax.

\cite{bailer-jones_estimating_2018} provides bayesian inferences of distances for \textit{Gaia} DR2 sources using a distance prior that assumes an exponential decrease in the space density of sources in the line of sight, with a length scale that depends on the location in the plane of the sky. This method allows to retrieve a distance estimate even for sources with negative parallaxes. Their method returns a distance for IGR J16318-4848 of 5.2$_{-1.8}^{+2.7}$\,kpc, using a length scale of 1.375\,kpc. This result is compatible with locating IGR J16318-4848 close to the SFR [2] discussed earlier, at around 4.9\,kpc.

The prior was chosen so that it can produce a consistent catalogue of distances over all the \textit{Gaia} sources. However, the authors argue that any additionnal relevant constraint on the distance of a specific source can be used to refine the estimation. We suggest that we can use the two SFRs discussed earlier to produce a custom prior for IGR J16318-4848, and compute a more representative distance estimate.

\begin{figure}[h]
\fig{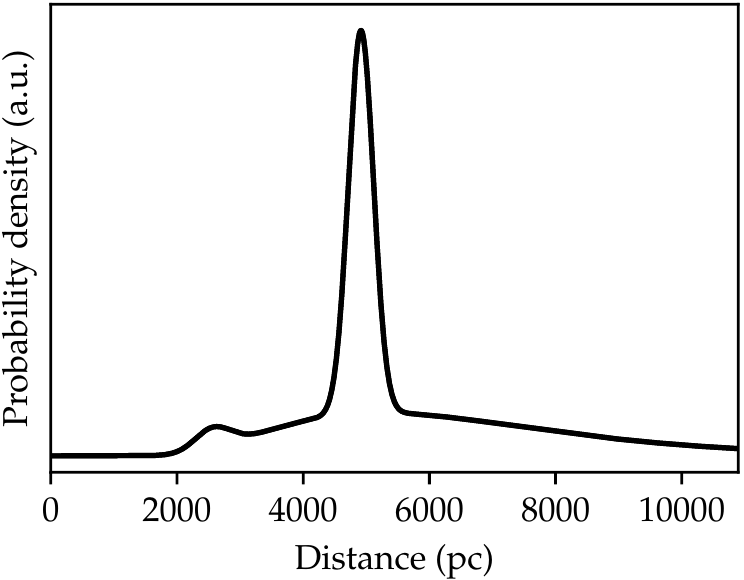}{.45\textwidth}{}
\caption{Distance probability density with custom prior for IGR J16318-4848.\label{fig:posterior}}
\end{figure}

We base ourselves on the prior used in \cite{bailer-jones_estimating_2018} and use the same length scale as they did for IGR J16318-4848 (1.375\,kpc). On top of that, we add two gaussian priors corresponding to the two SFRs in the line of sight, weighted by their separation to IGR J16318-4848 in the plane of the sky. We produce a posterior distribution of the distance probability density shown in \figref{fig:posterior}.

The negative \textit{Gaia} parallax measurement makes it very unlikely that the source is located below 3\,kpc, even with the added SFR prior at 2.4\,kpc. The integration of the distribution around its maximum returns a slightly more constrained distance estimation of 4.9$_{-1.5}^{+1.9}$\,kpc.

\subsection{X-ray luminosity}\label{subsect:X-ray_lum}
\cite{iyer_orbital_2017} compiled the X-ray fluxes (4--11\,keV) of IGR J16318-4848 through its orbital phases, obtained by \textit{XMM-Newton}, \textit{Swift}, \textit{ASCA}, \textit{NuSTAR} and \textit{Suzaku}, each fitted by an absorbed power law with an extra Fe-K$\alpha$ emission line. The 4--11\,keV flux varies from 2.29 to 11.59\tentothe{-12}\,erg\,cm$^{-2}$\,s$^{-1}$. For an isotropic emission, that would correspond to an X-ray luminosity of 1.5--8.0\tentothe{33}\,erg\,s$^{-1}$ at 2.4\,kpc,  0.7--3.3\tentothe{34}\,erg\,s$^{-1}$ at 4.9\,kpc, and up to 1.2--6.4\tentothe{34}\,erg\,s$^{-1}$ at 6.8\,kpc.

In all of the above cases, this puts IGR J16318-4848 towards the category of low-luminosity HMXBs (L$_X<$ 4\tentothe{36}\,erg\,s$^{-1}$), and is compatible with a wind-fed system.

\begin{figure*}[ht]
\fig{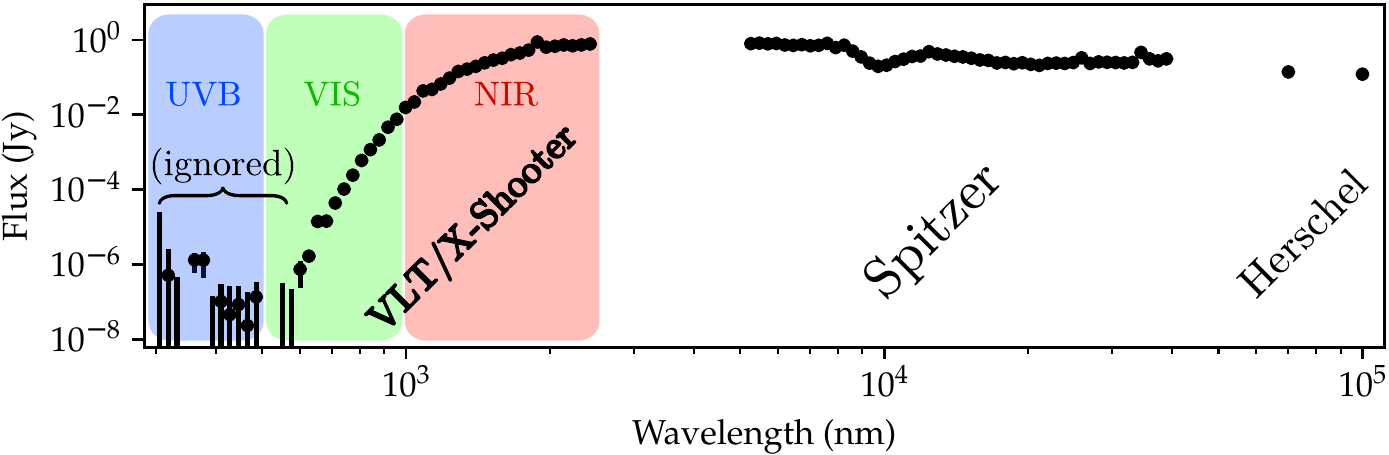}{.9\textwidth}{}
\caption{Broadband SED of IGR J16318-4848 obtained from the X-Shooter, \textit{Spitzer} and \textit{Herschel} data.\label{fig:sed_raw}}
\end{figure*}

\section{Fitting the spectral energy distribution\label{sect:SED}}

In this section, we model the SED of IGR J16318-4848 using the various results from spectroscopy with X-Shooter in the scope of an sgB[e] donor star surrounded by a circumstellar disk. However, we note that the spectral type of the companion was never directly tested, neither by its effective temperature (via H$\beta$ for instance) nor by stellar atmosphere modeling. For instance, \cite{filliatre_optical/near-infrared_2004} argue that a Of/WN companion should not display as much hydrogen, however a cool WN star may have a hydrogen mass fraction up to 50--60\% (\citel{sander_Wolf-Rayet_2014}, \citel{hainich_Wolf-Rayet_2014}). Moreover, if the P-Cygni lines form in the dense stellar wind, the wind would produce a non-blackbody continuum that may contribute significantly to the SED.

In this section, we thus work with the sgB[e] companion hypothesis, and we briefly test the dense wind donor case in Sect.\,\ref{sect:power}.

\subsection{The broadband dataset}
To complete the X-Shooter data, we retrieved archival \textit{Spitzer} spectra and \textit{Herschel} data to build the broadband spectral energy distribution (SED) from optical to mid-infrared. Because the spectra from X-Shooter and \textit{Spitzer} count more than 100\,000 spectral bins in total and we do not fit any synthetic stellar spectra, we binned the data in order to reduce the computing time and smooth out the spectral features. Since the full SED spans over 2-3 orders of magnitude in wavelength, we chose to evenly sample the X-Shooter and \textit{Spitzer} data in the logarithmic space, so that we end up with 50 binned data points for each of the two spectra. We did not bin the \textit{Herschel} data since we only have two data points, at 70 and 100\micron. This binning of the X-Shooter spectrum allows us to partially overcome the high reddening and recover the signal of the continuum down to 600\,nm, as presented in \figref{fig:sed_raw}. In the following, all fits ignore the data below this wavelength. We assume that the orbit of the system is circularized, and that the compact object is a 1.4\Msun\, neutron star.

\subsection{Source model, geometry and absorption}

\subsubsection{Dusty disk contribution}
We use the 2D flat model presented in \cite{lachaume_resolving_2007} for the dusty disk, along with the temperature evolution law across the radius T$_{disk}$(r) at a fixed temperature index $q$:

\begin{equation}\label{eq:disktemp}
T_{disk}(r) = T_{in}\left(\frac{r}{R_{in}} \right)^{-q}
\end{equation}

Its inner radius is set at \Rrim, where the temperature is T$_{disk}$(\Rrim) = \Tin. Its total contribution is the sum of the flux radiated by annuli of radius $r$ and width $dr$ at the temperature T$_{disk}$(r):

\begin{equation}\label{eq:diskflux}
F_{\nu, disk} = \frac{2\pi~cos(i)}{D^2}~\int_{R_{in}}^{R_{out}} rB_{\nu}(\nu, T(r))~dr
\end{equation}

Following the results in \cite{chaty_broadband_2012}, we fixed $q=0.75$ which corresponds to a viscous disk.

\subsubsection{Irradiated rim contribution}
We consider the inner rim to be a portion of cylinder of uniform temperature \Trim. Its effective area depends heavily on the viewing angle, its radius \Rrim\, and half-height \Hrim\, because of self-occultation; it is calculated using the formula given in \cite{dullemond_passive_2001}. The authors define the parameter $\delta = tan(i)\times H_{rim}/R_{rim}$; if $\delta >1$ (high inclination, high \Hrim\,to \Rrim\, ratio), the surface area is given in Eq.\,\ref{eq:rimsurface1}, and if $\delta \leq 1$ (lower inclination, low high \Hrim\,to \Rrim\, ratio), the surface area is given by Eq.\,\ref{eq:rimsurface2}

\begin{equation}\label{eq:rimsurface1}
S_{rim} = 2R_{rim}^2 cos(i)[\delta \sqrt{1-\delta^2} + arcsin(\delta)]
\end{equation}

\begin{equation}\label{eq:rimsurface2}
S_{rim} = \pi R_{rim}^2 cos(i)
\end{equation}

The total contribution of the irradiated rim is thus given by equation \ref{eq:rimflux}.
\begin{equation}\label{eq:rimflux}
F_{\nu, rim} = \frac{S_{rim}}{D^2}~B_{\nu}(\nu, T_{rim})
\end{equation}

\subsubsection{Stellar contribution}
The central star of radius \Rstar\, has a fixed temperature of \Tstar\,=20\,000\,K, following the same assumption made in \cite{chaty_broadband_2012}. Its effective area is calculated taking into account the occultation by the rim.

The effective area is obtained considering two different cases, depending on which of \Rstar\, or \Hrim\, is larger. For readability, we use $\alpha = \frac{\pi}{2} - i$ in the following equations. If \Rstar $<$ \Hrim, there are three cases to consider. The star is fully visible (low inclination), fully occulted (high inclination) or partially occulted by the rim. Depending on the relative size of the star and the disc, we find the following specific angles:

\begin{eqnarray}
\alpha_m = 2\,arctan\left( \frac{-R_{rim} + \sqrt{H_{rim}^2 + R_{rim}^2 - R_*^2}}{R_* + H_{rim}} \right) \\
\nonumber
\alpha_M = 2\,arctan\left( \frac{R_{rim} - \sqrt{H_{rim}^2 + R_{rim}^2 - R_*^2}}{R_* - H_{rim}} \right)
\end{eqnarray}

If $\alpha \geq \alpha_M$, the star is fully visible and its effective surface is a disk. If $\alpha  \leq \alpha_m$, the star is fully occulted by the rim. If $\alpha_m < \alpha < \alpha_M$, the star is partially occulted and its effective surface S$^*$ is determined by:

\begin{eqnarray}\label{eq:starsurface}
\nonumber
&h(\alpha) = R^* - cos(\alpha)\,(H_{rim} - R_{rim}\,tan(\alpha)) \\
\\
\nonumber
&S^* = R^{*2}\,arccos\left(1 - \frac{h(\alpha)}{R^*}\right) - (R^* - h)\sqrt{2\,R^*h(\alpha) - h^2(\alpha)}
\end{eqnarray}

If \Rstar $>$ \Hrim, the central star is either fully visible if $\alpha > \alpha_M$ or partially occulted by the rim and the disk if $\alpha < \alpha_M$. In the latter case, the effective area is again obtained with Eq.\,\ref{eq:starsurface}. The stellar contribution is:

\begin{equation}\label{eq:starflux}
F_{\nu}^* = \frac{S^*}{D^2}~B_{\nu}(\nu, T^*)
\end{equation}

\begin{figure*}

\fig{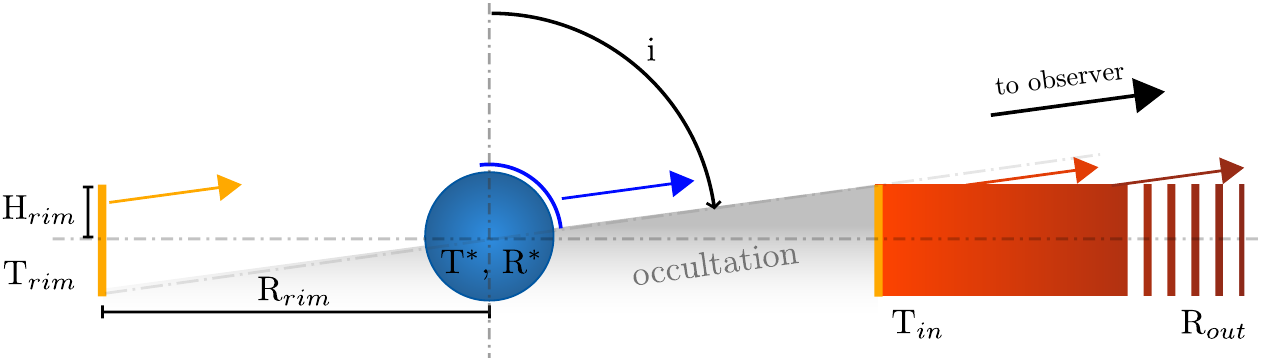}{.9\textwidth}{}
\caption{Edge-on view of the adopted geometry for the fit of the SED. The central star is in blue, the irradiated rim in orange and the dusty disk in red. The disk, which probably extends much further to scale, is only shown truncated on the right side for the sake of readability.\label{fig:sedmodel}}
\vspace{0.75cm}
\end{figure*}

A complete summary of the adopted geometry is shown in Fig.\,\ref{fig:sedmodel}.

\subsubsection{Absorption in the line of sight}
The latest measurement of the absorption towards IGR J16318-4848 comes from \cite{chaty_broadband_2012} at A$_V$=18.3. As discussed in Sect.\,\ref{subsect:dibs}, we are no able to update that value with our current data. To fit the data, all the source functions will be reddened using the formula from \cite{cardelli_relationship_1989} for optical/nIR and from \cite{chiar_pixie_2006} for mIR, taking into acount the silicate absorption features at 9.7 and 18\,\micron.

\subsection{Different aproaches for SED fitting}
For the sake of completion, we explore the two cases for which IGR J16318-4848 is located at either D = 2.4 or 4.9\,kpc, corresponding to the SFRs we discuss in Sect.\,\ref{subsect:SFR}. The model we use to fit the SED is then governed by the following parameters: \Rstar, \Tstar, \Rrim, \Hrim, \Trim, \Rout, \Tin, i, D and A$_V$. We perform the fit with D and A$_V$ fixed. Even then, the problem is highly degenerated and we thus use additional constraints for the fit to converge, which we discuss in the following sections.

\subsubsection{Fixing \Trim\, at the dust sublimation temperature}
In this section, the rim emission is constrained by the central star. Its temperature is fixed at 1\,500\,K, the dust sublimation temperature. Its covering factor \Hrim/\Rrim\ (H/R hereafter), is a variable and its radius is computed via conservation of stellar flux to match the rim temperature.

In this case, the fit converges towards a star of radius 38\Rsunperkpc\, independently of the distance chosen, with a temperature of 6\,600\p 300\kelvin. The rim covering factor H/R converges towards zero, suggesting the rim has close to zero contribution in the flux. The dusty disk is poorly constrained, with an inner temperature of 800\p 400\kelvin\, and an outer radius of 80\p 70\auperkpc. The inclination angle is 89.4\p 0.9\degr, which is degenerated as it is compatible with the disc both contributing to the SED and being completely invisible.

The stellar parameters are by themselves not compatible with the sgB[e] hypothesis, as the stellar temperature is too cool to match an early supergiant star. The absence of any rim contribution is also in contradiction with the results on spectroscopy. This suggests that the central star does not dominate the emission in IGR J16318-4848.

\subsubsection{Adding extra constraints from spectroscopy}
In this section, we make the hypothesis that the rim dominates the SED in nIR. The parameters of the star are computed from the fitted values of the rim. The equation from \cite{dullemond_passive_2001} that we use to compute the stellar radius is the following:

\begin{equation}
R^* = R_{rim}\left(\frac{T_{rim}}{T^*}\right)^2 \left(1+\frac{H_{rim}}{R_{rim}}\right)^{-\frac{1}{2}}
\end{equation}

It corresponds to the conservation of radiative flux from the star to the rim, assuming the entire flux received by the rim is used to heat it. The last term in the equation takes into account the self-irradiation of the rim, which becomes more important as its height becomes larger.

The stellar temperature is fixed at 20\,000\kelvin\, as suggested in previous studies \citep{filliatre_optical/near-infrared_2004,chaty_broadband_2012}. We also note that this temperature is compatible with the models from \cite{vink_fast_2018} that predict slow wind velocities (v\,$\lesssim$\,500\kmpersec) for supergiant stars cooler than 21\,000\,K, which is the case for the wind of IGR J16318-4848.

The stellar radius will be determined by the fitted temperature, radius and width of the rim. In this case, the rim contribution is degenerated since three parameters influence its effective area of emission. Thus, we use spectroscopy results on the orbital velocity of the rim to constrain its absolute radius assuming a circular keplerian orbit.

We retrieve the average projected orbital velocity v$_r$sin(i)=113\kmpersec\, from the double-peaked profiles identified in Sect.\,\ref{subsect:Hmodel}. Given a central mass (i.e. the sellar mass plus the compact object), we can derive a spectroscopic radius of the rim for a certain inclination angle. If we make the hypothesis that the compact object orbits within the cavity, it provides a minimum central mass so that the spectroscopic rim radius is higher than the compact object orbit:
\begin{equation}
M_{min} = \frac{P v_r^3sin^3(i)}{2\pi G}
\end{equation}
with P the orbital period of the compact object and v$_r$ the orbital velocity of the rim. In the extreme case of a purely edge-on view, the rim orbital velocity of 113\kmpersec\, gives an absolute minimal central mass of 12\Msun. Below that, the rim orbits closer to the star than the compact object.

Then, for higher central masses, the lowest inclination angle that is compatible with the previous orbit considerations is:

\begin{equation}
sin(i_{min}) = v_r \left( \frac{P}{2\pi GM} \right)^{\frac{1}{3}}
\end{equation}

The central mass should not be much higher than 50\Msun. For the compact object to orbit whithin the cavity, the rim orbital velocity measurement implies a minimum inclination angle of i$_{min}$ = 49\degr. As IGR J16318-4848 is expected to be wind-fed, this provides a lower limit on the inclination during the fit.

\subsubsection{Results}\label{subsubsect:sedresults}

\begin{figure}[h]

\fig{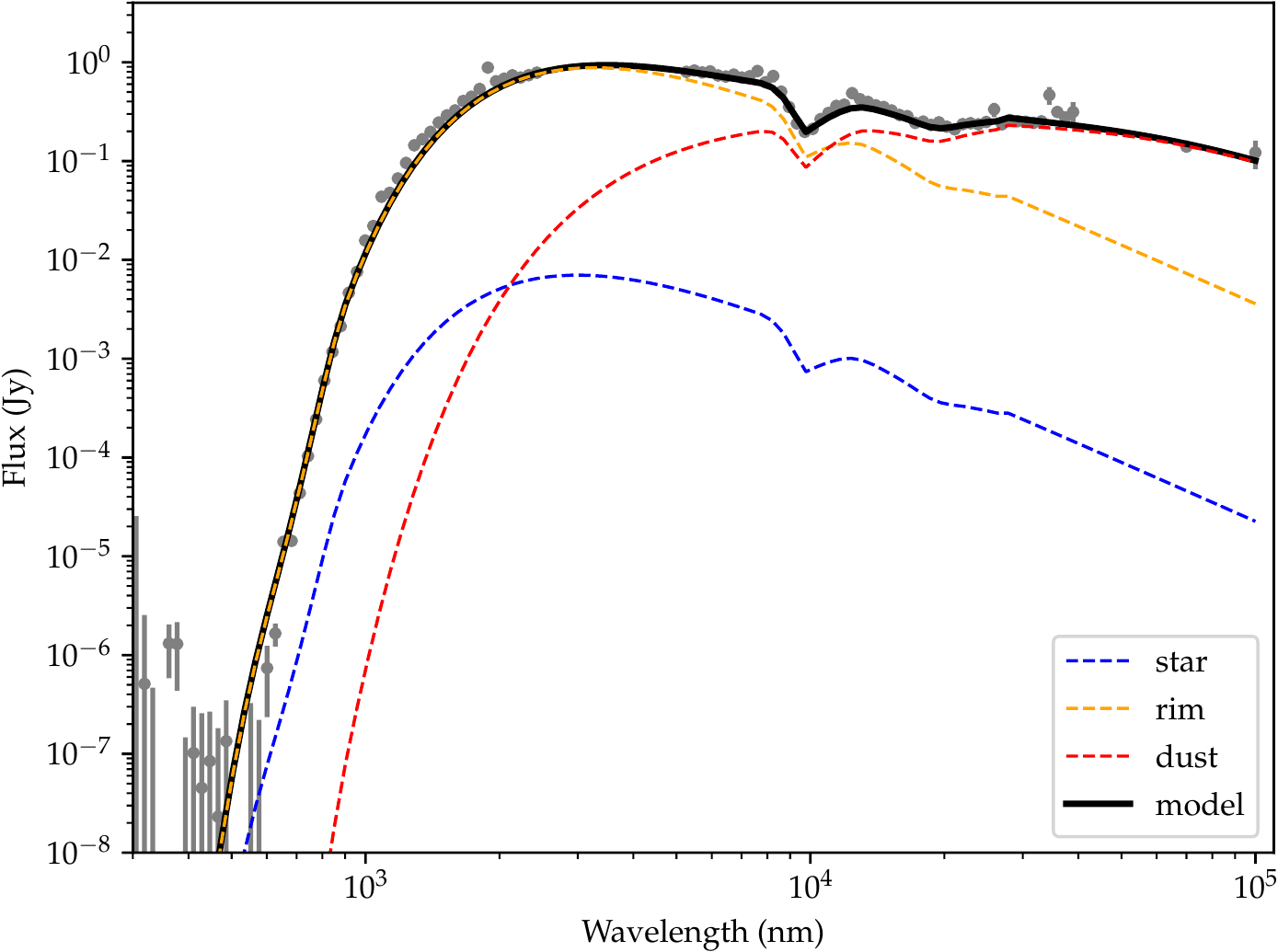}{.45\textwidth}{(a)}

\fig{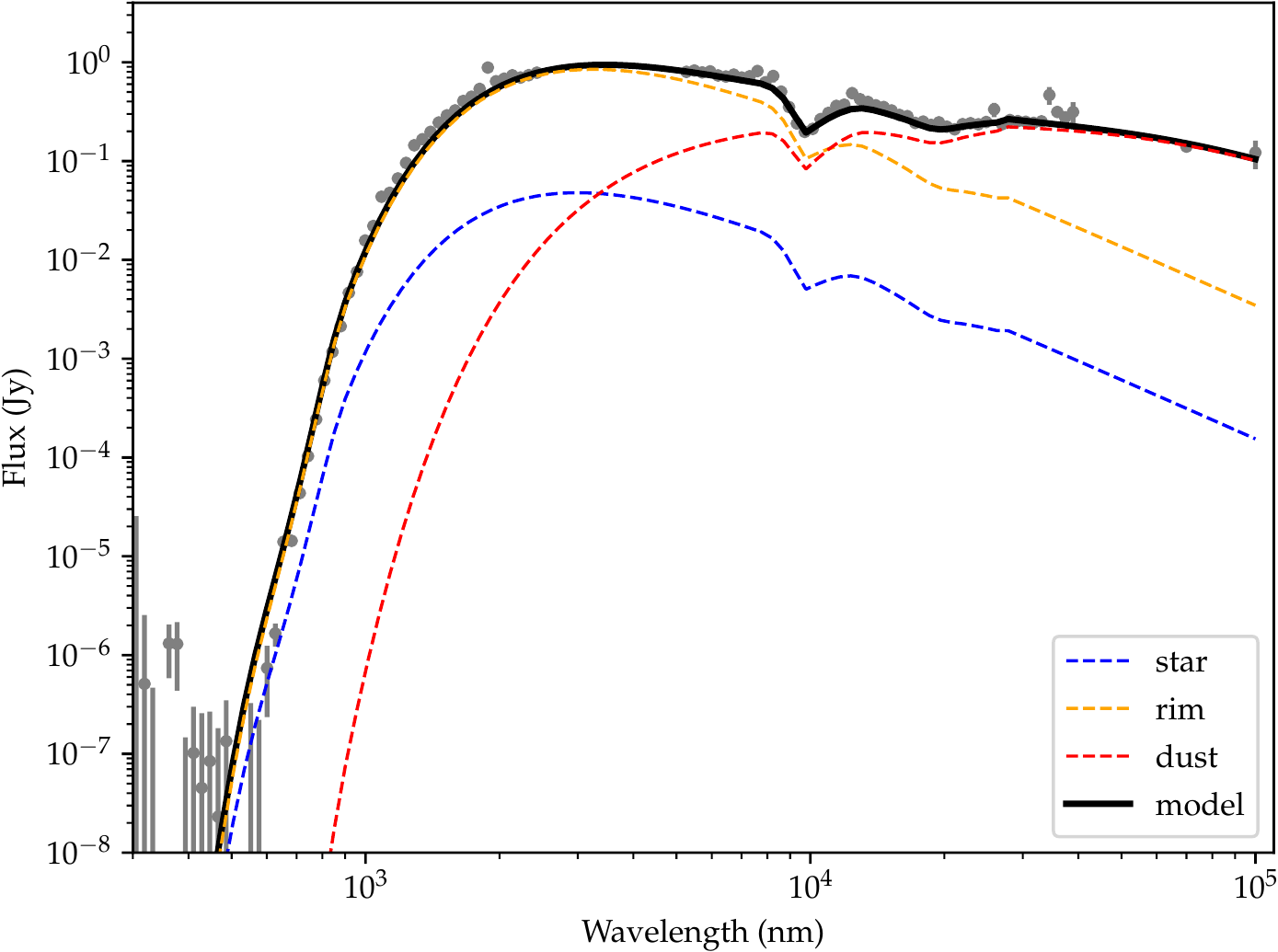}{.45\textwidth}{(b)}
\caption{Results of the SED fit for a distance of 2.4\,kpc, with a central mass of (a): 25\Msun\, and (b): 50\Msun.\label{fig:sed:24}}
\end{figure}

\begin{figure}[h]
\fig{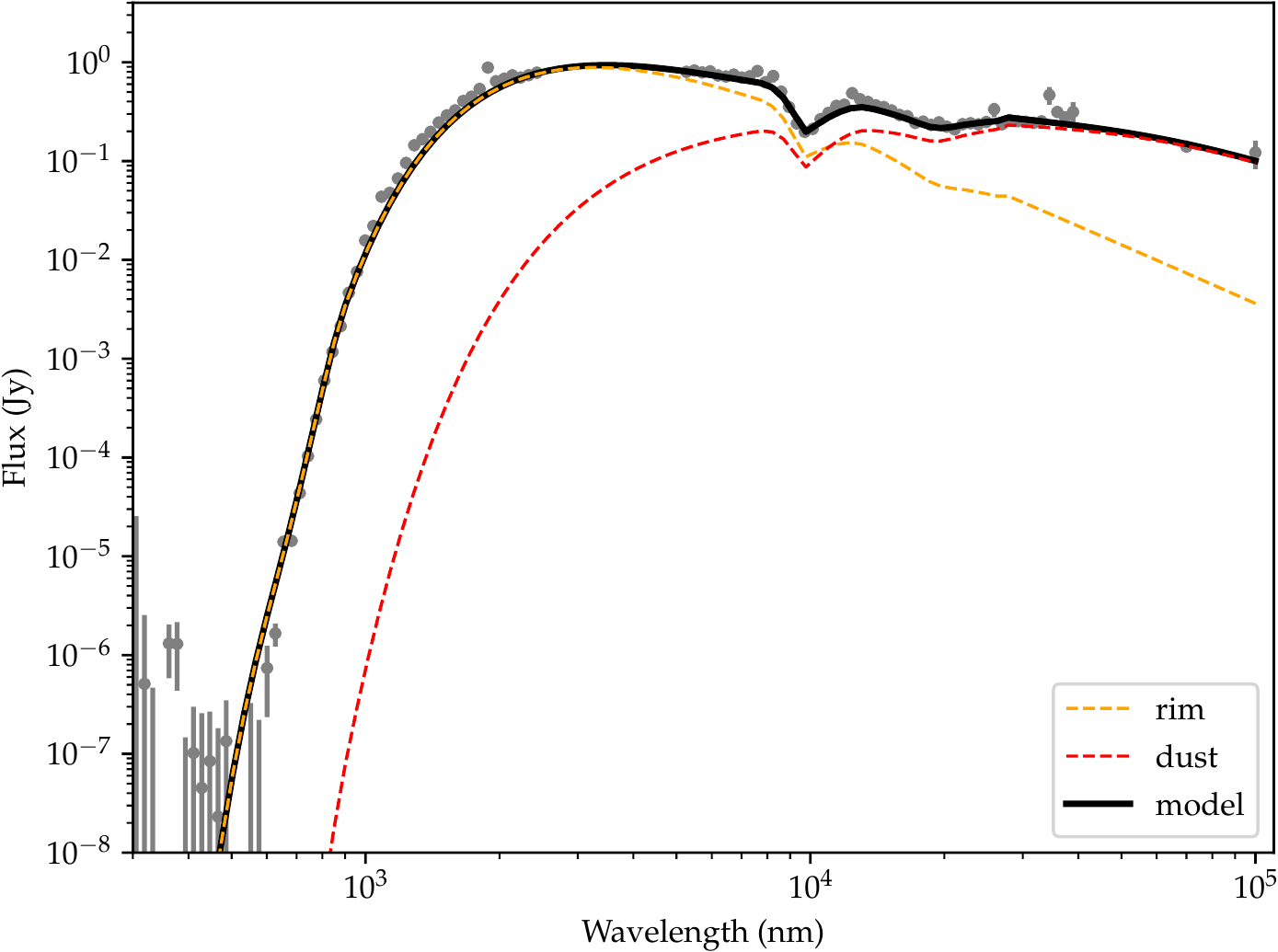}{.45\textwidth}{(a)}

\fig{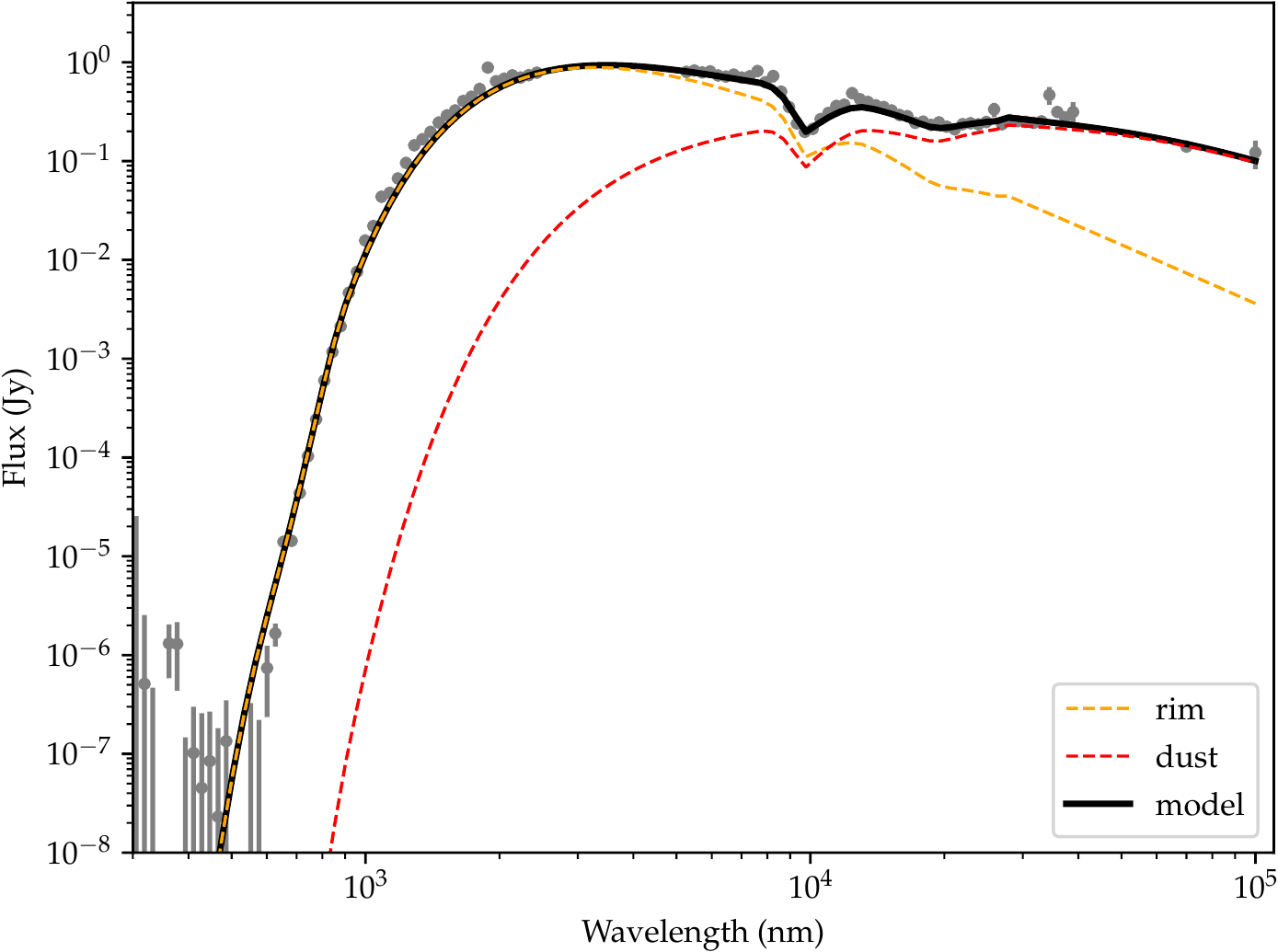}{.45\textwidth}{(b)}
\caption{Results of the SED fit for a distance of 4.9\,kpc, with a centra mass of (a): 25\Msun\, and (b): 50\Msun.\label{fig:sed:49}}
\end{figure}

The numerical results of the fit are presented in \tabref{tab:sed} for various distance value, spaning from 2.4 to 6.8\,kpc. We illustrate the modeled SEDs in  \figref{fig:sed:24} and \ref{fig:sed:49} for 2.4 and 4.9\,kpc only, as the 3.4 and 6.8\,kpc cases cannot be distinguished from the 4.9\,kpc SED.

The primary result comes from the fact that all the fits we performed converge to a rim temperature of \Trim\, = 6740\p 210\kelvin, and an inner disk temperature of \Tin\,=1374\p 47\kelvin.

Considering the period of 80.09\p 0.01\,d, the orbit of the compact object ranges from 1.06\,au (at \Mstar = 25\Msun) to 1.34\,au (50\Msun). From our fit results, this places the compact object within the cavity inbetween the central star and the rim. This further supports, along with the low X-ray luminosity (sect. \ref{subsect:X-ray_lum}), that IGR J16318-4848 is a wind-fed system.

We note that for a source located at 2.4\,kpc, the fit converges towards an inclination angle that is off from an edge-on view only by a few degrees, in compatibility with our assumption that \Halpha\, comes from the polar wind of the central star (Sect.\,\ref{subsect:inclination}). However, this configuration allows a geometry that makes the central star visible and emit a continuum that is a hundred to a thousand times fainter than the continuum of the irradiated rim at 1\micron.

As for the case of 4.9\,kpc, the inclination can deviate significantly from an edge-on configuration if the central mass is too low; however this implies a greater height for the rim, which makes the ratio H/R grow outside the range of expected values (0.1--0.3, \citel{dullemond_passive_2001}). Thus, we suggest that a reasonable lower limit for the inclination of IGR J16318-4848 is i$_{min}$=76\degr, which is obtained when fitting a 25\Msun\, star at 4.9\,kpc. In this configuration, the geometry of the circumbinary material does not allow the star to be visible.

For the furthest distance estimate of 6.8\,kpc, the fit converges towards very high values of H/R ($\geq$ 0.8) for a 25\Msun\, star and the rest of the parameters are not properly constrained.

Overall, the SED modeling tends to favor higher central masses because it would otherwise imply a too large H/R ratio, except if the source is close enough. We reckon that a lower limit for the inclination is 76\degr, but the best-fitting cases return a higher inclination in the range 86--88\degr, which agrees with the conclusion we draw from the polar wind \Halpha\, line. We note that in all cases, the inclination is high enough to allow the compact object to orbit within the cavity. This further supports that IGR J16318-4848 is a wind-fed system.

\vspace{1cm}

IGR J16318-4848 is one of the most absorbed of the supergiant B[e] X-ray binaries in the Galaxy. It has a very complex and dynamical environment that is challenging to observe and interpret. While we managed to draw a clearer picture of its local medium, there is yet to have an unambiguous explanation for the origin of the circumbinary material. It is one of the many exotic features of this source, and it could be either due to the natural evolution of the massive companion or result from binary interaction.

\section{Modeling the stellar atmosphere and wind} \label{sect:power}
\subsection{The PoWR model}

As an alternative approach to the simple 20\,000\,K sgB[e] companion with a minor contribution presented in Sect.\,\ref{sect:SED}, we test here the assumption of a major contribution from a massive donor with the same temperature and an optically thick wind responsible for the P-Cygni lines. The stellar atmosphere modeling was performed using the Potsdam Wolf-Rayet (PoWR) model atmosphere code. This code solves the comoving-frame raditive transfer together with the solution of the population numbers and the temperature stratification in full non-LTE (e.g. \citel{hamann_temperature_2003}). It fully accounts for iron-line blanketing \citep{grafener_line-blanketed_2002} and optically thin densitiy inhomogeneities ("microclumping", \citel{hamann_spectrum_1998}). The quasi-hydrostatic part is treated self-consistently \citep{sander_consistent_2015}, while a $\beta$-law is assumed in the supersonic domain. The PoWR model provides a stellar atmosphere stratification as well as an emergent spectrum over a wide wavelength range, from the (E)UV to the mid-IR.

\subsection{Results of the modeling}
\begin{figure}
\begin{minipage}{.4\textwidth}
\fig{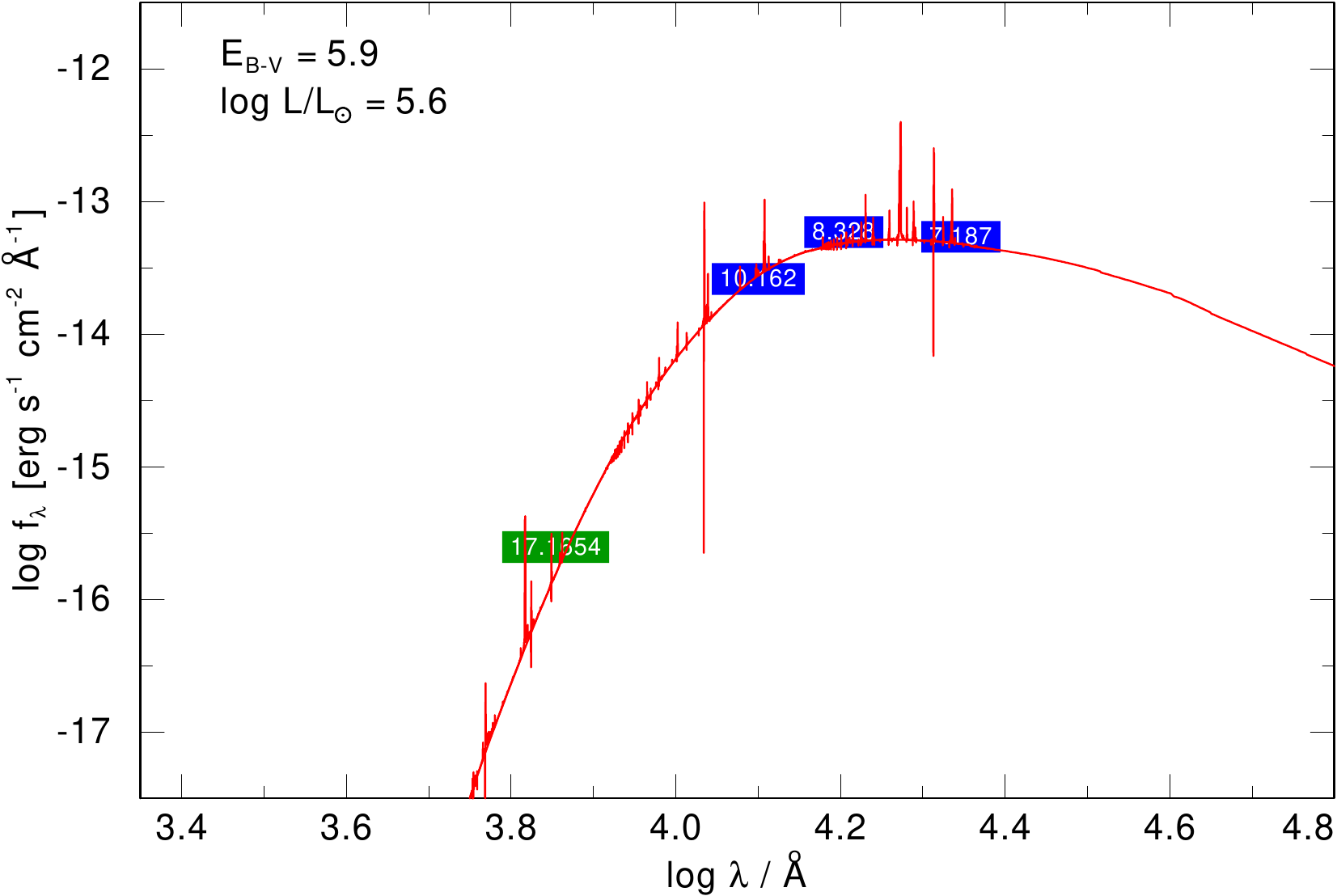}{\textwidth}{}
\caption{Comparison of the computed SED from the PoWR code with available photometric data, assuming a 2.4\,kpc distance.\label{fig:model_SED}}
\end{minipage}
\hfill\vspace{0.2cm}
\begin{minipage}{.4\textwidth}
\fig{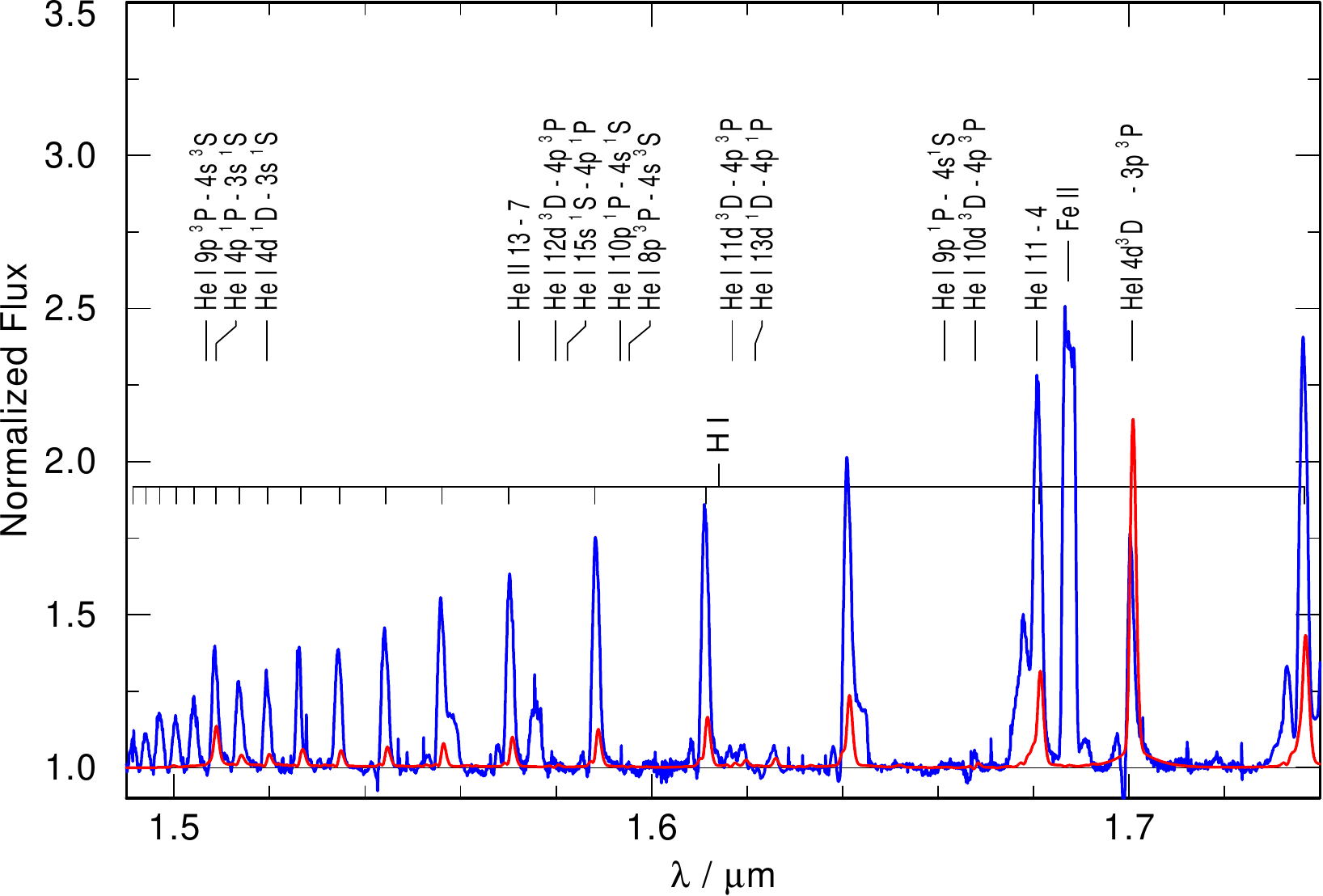}{\textwidth}{}
\caption{Hydrogen lines reconstructed from the PoWR model (red) versus the normalized input spectrum (blue).\label{fig:model_hlines}}
\end{minipage}
\end{figure}

\begin{figure}
\fig{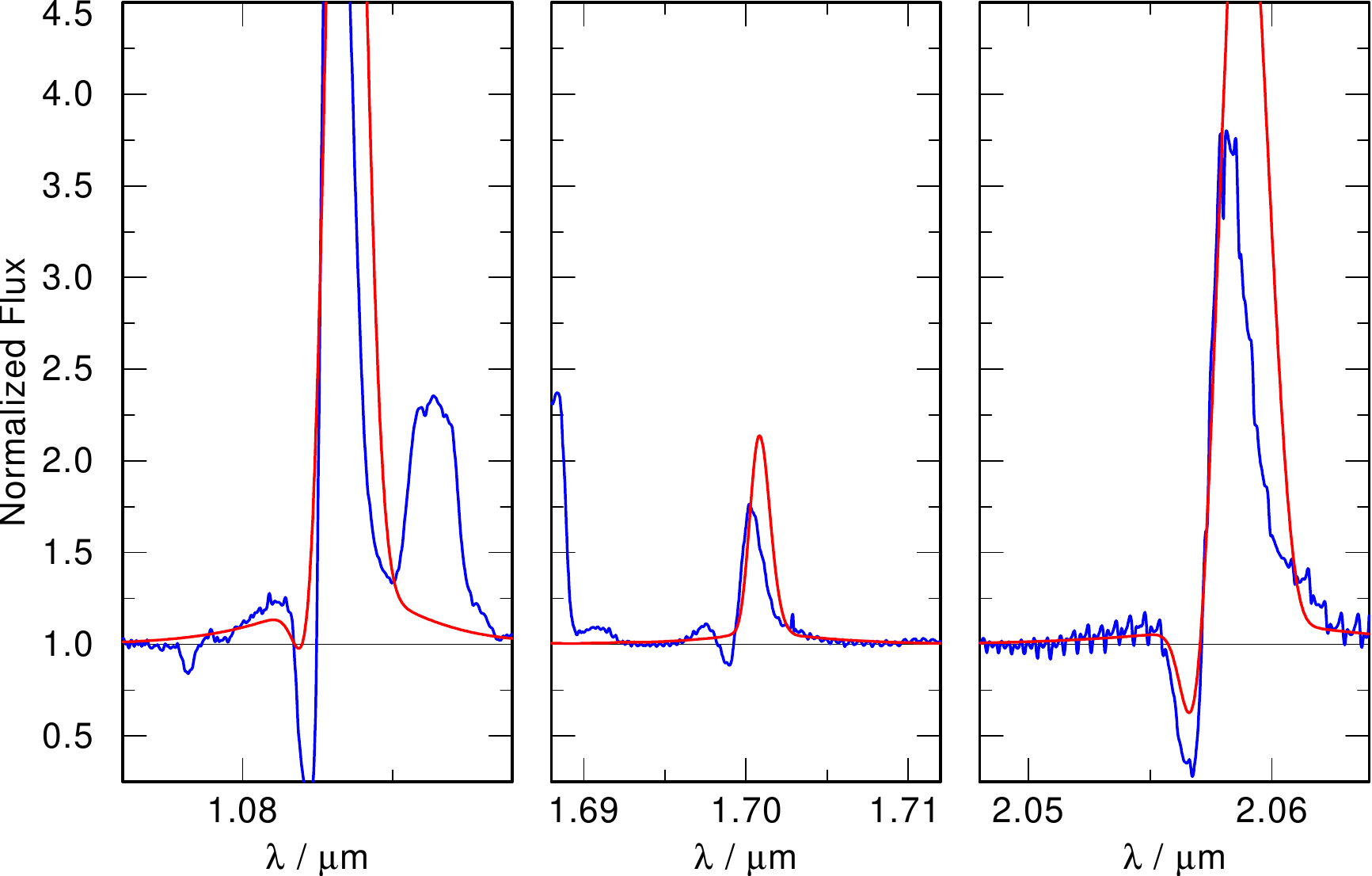}{.4\textwidth}{v$_{\infty}$ = 300\kmpersec}\label{fig:windlines300}

\fig{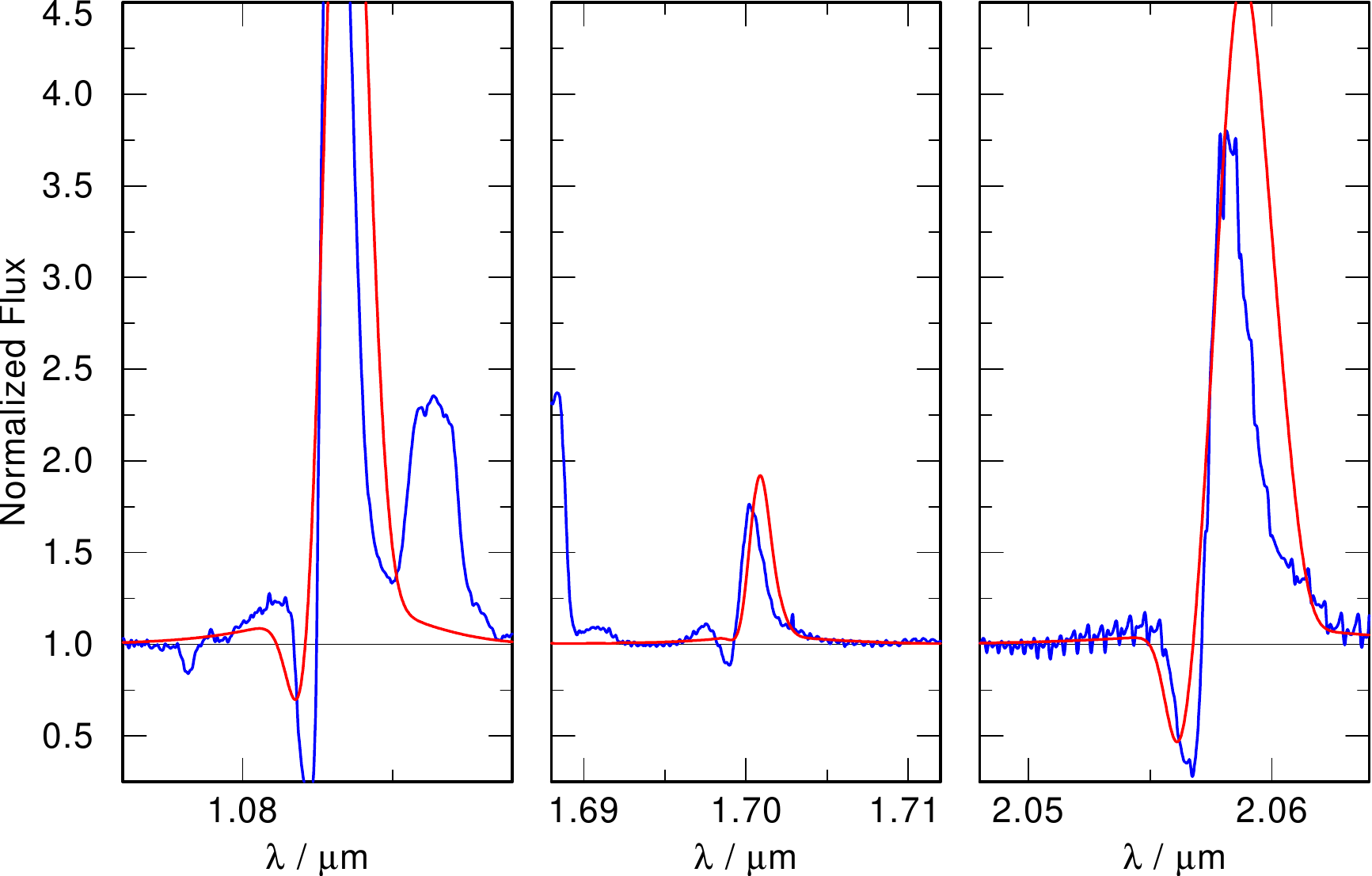}{.4\textwidth}{v$_{\infty}$ = 400\kmpersec}\label{fig:windlines400}
\caption{Wind lines reconstructed from the PoWR model for prominent helium transitions with different terminal velocities (red) versus the normalized input spectrum (blue).\label{fig:model_windlines}}
\end{figure}

In our Fig.\,\ref{fig:model_SED} we show the overall spectral energy distribution (SED) obtained with the model compared to the available \textit{Gaia} and JHK photometry assuming a distance set to the minimal value we derive of 2.4\,kpc. Moreover, Figs\,\ref{fig:model_hlines} and \ref{fig:model_windlines} highlight selected parts of the normalized model spectrum compared to the available observation to illustrate the contribution of the star to the hydrogen emission lines and prominent P-Cygni lines formed in the stellar wind. We note that we were not able to reproduce the flat-topped \ion{Fe}{2} lines. In Tab.\,\ref{tab:power_params} we present the results for the stellar and wind parameters. For these, we assumed a fixed effective temperature of \Tstar$ = 20\,500$\,kK at a Rosseland continuum optical depth of $20$, a hydrogen mass fraction in line with Sect.\,\ref{subsect:HHe} as well as a radius and mass combination from the constrains in Sect.\,\ref{subsubsect:sedresults} yielding a result closest to globally consistent energy budget.

The PoWR model results show P-Cygni profiles for hydrogen lines, however we are unable to reproduce their intensity as shown in Fig.\,\ref{fig:model_hlines}. We do recover the intensity for the helium emission lines, but the modeled P-Cygni absorption is usually weaker than the observations (Fig\,\ref{fig:model_windlines}). However, we did not expect to fully reproduce the line shapes as this would require a full decomposition into a stellar and a disc component, which is beyond the scope of the present paper. According to these results, we infer that the terminal velociy of the wind can be reasonably constrained to be between 300 and 400\kmpersec.

\subsection{Discussion}
In this first approach at SED modeling of IGR J16318-4848 with the PoWR code, we consider that the radiation mainly comes from the star and its close environment. We reckon that it is a sensible alternative to the minor donor contribution considered in Sect.\,\ref{sect:SED} and that it brings interesting results especially concerning the reproduction of line profiles.

The terminal velocity of the wind inferred with the model atmosphere is in agreement with the one derived in Sect.\,\ref{subsect:Hmodel}. Concerning the hydrogen lines, the fact that we are not able to repoduce them correctly supports our findings in sect.\,\ref{subsect:Hmodel} that there is considerable contribution from the irradiated rim.

We suggest that the hydrogen lines produced by the PoWR code can be attributed to the extra central line discussed in Sect.\,\ref{subsect:Hmodel} for which we previously failed to identify the nature, and in fact come from the vicinity of the central star. Also, this would indicate that the line of sight allows us to see at least part of the central star. However, our multi-component SED modeling in Sect.\,\ref{sect:SED} does not allow us to see the star if it is located at 4.9\,kpc. This may suggest that the source is closer, or that our geometrical model is not entirely accurate.

As for the helium lines, the stellar atmosphere and wind modeling reproduce their overall profile rather well. The P-Cygni absorption is still slightly more difficult to fully recover, probably because of the complex circumbinary environment in the line of sight that is not taken into account during the modeling (i.e. the dusty disc). Most of the helium lines are still much better reproduced than hydrogen in terms of intensity, width and P-Cygni absorption with the PoWR code. This result might indicate that the central star has blown off part of its hydrogen envelope through intense stellar wind. As such, the photosphere might be enhanced in helium and the circumbinary medium might have been partly formed from that wind, hence the intense hydrogen emission lines from the irradiated rim. Following the discussion in Sect.\,\ref{subsect:HHe}, this rather suggest an intrinsic origin for the enhanced He/H ratio.

Despite calculating outwards to 10$^5$\,\Rstar, the PoWR model could not reproduce any of the flat-topped \ion{Fe}{2} lines. While this would support the hypothesis that these transitions do not arise from the star nor its wind but from the circumbinary disk, their absence in the synthetic spectrum may also be due to the different treatment of the large amount of iron lines compared to other elements in PoWR. This will need to be investigated in a separate project.

The mass and radius of the central star recovered from the PoWR code are compatible with the SED fitting. However, the mass is on the lower end of our estimations, and our fit at 25\Msun\, converges towards a rim height to rim radius ratio that is higher than the expected range. The effective temperature of the photosphere T$_{eff}^*$ of 20\,500\,K is motivated by the value of 20\,000\,K suggested in \cite{filliatre_optical/near-infrared_2004} and \cite{chaty_broadband_2012}.

This first attempt at modeling the spectrum of IGR J16318-4848 hints that the exact nature of its donor is still up to debate, and that a dense wind donor (a late hydrogen-rich Of/WN) is a possibility. To clear it up, we reckon that further modeling with a complete parameter space exploration along with the subtraction of potential disk contribution is required.

\begin{table*}
\begin{minipage}[t]{.6\textwidth}
\caption{Results of the fit of the SED, with the rim radius constrained by spectroscopy.\label{tab:sed}}
\centering
\begin{tabular}{ccllcc}
\hline\hline\\[-1.5ex]
Mass (\Msun) & H/R & \Rout (au) & i (deg) & \Rstar (\Rsun) & \Rrim (au) \\
\hline\\[-1.5ex]
\multicolumn{6}{c}{\textbf{2.4\,kpc}} \\
25 & 0.19\p 0.01 &  110\p 50   & 87.07\p 0.02    & 38.7\p 0.2 & 1.74\p 0.01 \\
50 & 0.11\p 0.01 &  200\p 100  & 89.298\p 0.005  & 80.5\p 0.4 & 3.47\p 0.01 \\
\hline\\[-1.5ex]
\multicolumn{6}{c}{\textbf{4.9\,kpc}} \\
25 & 0.40\p 0.02 & 100\p 50   & 76\p  1         & 33.8\p 0.1 & 1.64\p 0.01 \\
50 & 0.18\p 0.01 & 200\p 100  & 86.93\p 0.02    & 77.8\p 0.3 & 3.46\p 0.01  \\

\multicolumn{6}{c}{\textbf{3.4\,kpc}} \\
25 & 0.33\p 0.01 & 110\p 50  & 84.0\p 0.3   & 36.4\p 0.2 & 1.72\p 0.01  \\
50 & 0.27\p 0.01 & 200\p 100 & 88.53\p 0.08 & 75.4\p 0.3 & 3.47\p 0.01  \\

\multicolumn{6}{c}{\textbf{6.8\,kpc}} \\
25 & \multicolumn{5}{c}{\textit{converges to unrealistic parameters}} \\
50 & 0.22\p 0.01 & 200\p 100 & 83.97\p 0.04 & 75.8\p 0.5 &  3.44\p 0.01 \\
\hline
\end{tabular}
\\[1.5ex]
\Trim\, = 6700\p 200\kelvin, \Tin\,=1370\p 50\kelvin.
\end{minipage}
\hfill
\begin{minipage}[t]{.35\textwidth}
\caption{Parameters used to model the atmosphere and wind of IGR J16318-4848 with PoWR.\label{tab:power_params}}
\begin{tabular}{ll}
\hline\hline\\[-1.5ex]
T$_{eff}^*$ & 20\,500\,K \\
T$_{2/3}$   & 18\,200\,K \\
X$_H$       & 0.5        \\
R$^*$       & 51\Rsun    \\
M$^*$       & 25\Msun    \\
log$\left(L/L_{\odot}\right)$ & 5.6 \\
log$(\dot{M})$ & -4.8 \\
v$_{\infty}$ & 400\kmpersec \\
D$_{\infty}$ & 10 \\
v$_{mic}$ & 14\kmpersec \\
\hline
\end{tabular}
\end{minipage}
\end{table*}

\section{Comparison with  the sgB[\lowercase{e}] HMXB CI Cameleopardis}

The first sgB[e] HMXB to be identified is CI Cameleopardis (CI Cam hereafter), and it was extensively studied since its discovery \citep{smith_IAUC_1998}. While both systems show different behaviors, we reckon the comparison with IGR J16318-4848 is relevant as these binaries might be of very similar nature or correspond to different phases of the same evolutionary path. \cite{bartlett_CI_2019} use recent \textit{Swift}/XRT observations of CI Cam spanning $\sim$150\,d and suggest a variability timescale between 75--100\,d, the same order of magnitude as the orbital period of 80\,d derived for IGR J16318-4848 by \cite{iyer_orbital_2017}.
\cite{robinson_high-dispersion_2002} suggest the presence of a high-velocity outflow (1000--2500\kmpersec) associated to a polar outflow seen almost pole-on \citep{hynes_spectroscopic_2002}.
\cite{bartlett_CI_2019} derive a column density of 6\ttentothe{22}--2\ttentothe{24}\NH, which reaches the estimated column density values of IGR J16318-4848. However, the dust extinction in CI Cam only reaches up to A$_V$=4\,mag \citep{hynes_spectroscopic_2002}, much lower than the extinction for IGR J16318-4848 (A$_V$=18.3, \citel{chaty_broadband_2012}). It is possible that this difference only comes from the viewing angle of those systems, which is inferred to be almost pole-on for CI Cam and almost edge-on for IGR J16318. Both having $N_{\text{H}}$ values much higher than their optical interstellar absorption suggest both have an absorbing medium local to the accretion region, i.e. around the compact object. The much higher A$_V$ in IGR J16318-4848 compared to CI Cam would be due to the presence of the equatorial circumbinary disc in the line of sight. Another notable feature in both sgB[e] systems is the presence of \ion{Fe}{2} flat-topped lines in their optical to infrared spectrum. However, while these profiles are present in quiescence for IGR J16318-4848, they were only observed in CI Cam during its 1998 outburst; instead, in quiescence, the profiles were shown to be double-peaked (or with a central depression,  see \citel{hynes_spectroscopic_2002}). This could stress on a significant difference between the two binaries, as CI Cam only displays a spherically symmetric outflow during outburst (flat-topped lines) and returns to an axisymmetric state in quiescence (double-peaked lines), while IGR J16318-4848 has no record of such behavior as the iron outflow persists in spherical symmetry. Another difference lies in the projected velocity of the expanding medium, which was measured to be 32\kmpersec\, in CI Cam \citep{robinson_high-dispersion_2002} and 250\p 20\kmpersec\, in IGR J16318-4848 (this study), about an order of magnitude higher.

\section{Conclusion}
We presented spectrocopy performed on IGR J16318-4848, with unprecedented resolution and coverage towards the optical band. With the analysis of spectral features, and the modeling of the broadband SED using archival \textit{Spitzer} and \textit{Herschel} data, we obtain the following results:

\begin{enumerate}

\item The inclination of the system is higher than 76\degr, and reaches up to 86--88\degr\, for the best-fitting models.

\item Based on SFR associations and the second data release of \textit{Gaia}, we infer the distance to IGR J16318-4848 to be 4.9$^{+1.9}_{-1.5}$\,kpc.

\item We confirm the presence of P-Cygni profiles in \ion{H}{1} and \ion{He}{1} lines that likely probe the equatorial wind of the central star expanding at velocities up to 370\kmpersec.

\item \ion{H}{1} lines with sufficient SNR can be reproduced by a double-peaked profile originating from the orbital motion of the rim at v$_r$sin(i)=113\p 4\kmpersec.

\item We resolve the flat-topped profiles of previously identified \ion{Fe}{2} and [\ion{Fe}{2}] lines in emission. This indicates they originate from an optically thin medium undergoing spherical expansion at 250\p 20\kmpersec. If their wing broadening comes from orbital motion, it would locate the origin of the wind at 3.4--6.8\,au away from the central star, meaning the iron lines form in a disk wind from the dusty equatorial disk.

\item The \Halpha\, line has an extra narrow component which, given the high inclination, could be associated to a fast polar wind seen almost edge-on.

\item Optical forbidden lines from [\ion{O}{1}], [\ion{N}{2}] and [\ion{S}{2}] display narrow profiles that suggests they could either originate from far away in the dusty disk (120--260\,au, or 700--740\,\Rstar), or from the foreground nebular emission of a spiral arm.

\item In the case of a sgB[e] companion, the irradiated rim temperature is fitted to be \Trim\,=\,6\,740\p 210\,K, and the temperature of the inner viscous disk is \Tin\,=\,1\,374\p 47\,K. In this model, those structures outshine by far the central star, which contributes at best to 10\% in optical flux.

\item In the case the compact object is a 1.4\Msun\, neutron star, it is likely to orbit within the central cavity. Adding the low X-ray luminosity we derived, this would suggest the system is wind-fed, though we can not rule out tidal interactions between the compact object and the inner disc.

\item We perform stellar atmosphere and wind modeling of the X-Shooter optical to near-infrared spectrum with the PoWR code, assuming a donor with a dense wind. The results are overall in good agreement with the observations, but further modeling with complete parameter space exploration might shed a new light on the exact nature of the donor.

\item We compare IGR J16318-4848 with CI Cam and note both systems might be of similar nature, and that the differences in some of their key features could be explained by the very different viewing angle (edge-on for IGR J16318-4848, pole-on for CI Cam).

\end{enumerate}

\section*{Acknowledgments}
\begin{footnotesize}
This work was supported by the Centre National d'Etudes Spatiales (CNES), based on observations obtained with MINE --Multi-wavelength INTEGRAL NEtwork--.
SC is also grateful to the LabEx UnivEarthS for the funding of Interface project "Galactic binaries towards merging".
This publication makes use of VOSA, developed under the Spanish Virtual Observatory project supported from the Spanish MICINN through grant AyA2011-24052;
of NASA’s Astrophysics Data System Bibliographic Services, operated by the Smithsonian Astrophysical Observatory under NASA Cooperative Agreement NNX16AC86A;
of data from the European Space Agency (ESA) mission {\it Gaia} (\url{https://www.cosmos.esa.int/gaia}), processed by the {\it Gaia} Data Processing and Analysis Consortium (DPAC, \url{https://www.cosmos.esa.int/web/gaia/dpac/consortium}). Funding for the DPAC has been provided by national institutions, in particular the institutions participating in the {\it Gaia} Multilateral Agreement.
A.\,A.\,C.\, Sander would like to thank STFC for funding under grant number ST/R000565/1.
\end{footnotesize}

\newpage
\appendix

\section{List of identified transitions}
\begin{small}
Each table provide the reference wavelength in air ($\lambda_0$, nm), the measured heliocentric velocity of the line center (V$_{h}$, \kmpersec), the flux density of the line (Jy), the full width at half-maximum (FWHM, \kmpersec) and the equivalent width (EQW, nm). For \ion{H}{1} and \ion{He}{1}, we also provide the heliocentric velocity difference between the emission and absorption component of the line (V$_{PC}$, \kmpersec). Typical errors are velocity are 5\kmpersec\, in velocity and 0.002\,nm in equivalent width. Hydrogen lines in bold benefited from more accurate modeling, their set of parameters are available in Sect.\,\ref{subsect:Hmodel}.
\end{small}

\begin{table}[h]
\footnotesize
\begin{minipage}[t]{0.5\textwidth}
\centering
\vspace{0pt}
\caption{Identified \ion{H}{1} lines\label{tab:line:hydrogen}}
\centering
\begin{tabular}{llllllll}
\hline\hline\\[-1.5ex]
Line & $\lambda_0$ & V$_{h}$ & Flux & \textsc{fwhm}        & \textsc{eqw}  & V$_{PC}$     \\
\hline\\[-1.5ex]
\Halpha\, 3-2 & 656.279 & -73 & 2.31e$^{-4}$ & 383 & -27.317& 165 \\
Pa 23-3 & 834.554  & -43 & 1.34e$^{-4}$ & 224 & -0.097 & ...    \\ 
Pa 22-3 & 835.900  & -13 & 2.47e$^{-4}$ & 266 & -0.207 & ...    \\
Pa 21-3 & 837.448  &~~~5 & 2.57e$^{-4}$ & 299 & -0.264 & ...    \\
Pa 20-3 & 839.24   & -15 & 2.45e$^{-4}$ & 322 & -0.274 & ...    \\
Pa 19-3 & 841.332  & -17 & 3.99e$^{-4}$ & 274 & -0.358 & ...    \\
Pa 17-3 & 846.726  & -28 & 6.65e$^{-4}$ & 358 & -0.661 & 254  \\
Pa 16-3 & 850.249  & -37 & 5.82e$^{-4}$ & 262 & -0.373 & ...    \\
Pa 15-3 & 854.538  & -46 & 9.04e$^{-4}$ & 308 & -0.746 & 352  \\
Pa 14-3 & 859.839  & -37 & 7.87e$^{-4}$ & 275 & -0.454 & 272  \\
Pa 13-3 & 866.502  & -41 & 1.30e$^{-3}$ & 277 & -0.745 & 313  \\
Pa 12-3 & 875.046  & -48 & 1.49e$^{-3}$ & 356 & -0.921 & 224  \\
Pa 11-3 & 886.289  & -33 & 2.15e$^{-3}$ & 301 & -0.928 & 257  \\
Pa 10-3 & 901.533  & -83 & 3.17e$^{-3}$ & 377 & -1.182 & 202  \\
Pa 8-3  & 954.618  & -33 & 9.80e$^{-3}$ & 324 & -1.415 & 275  \\
Pa 7-3  & 1004.937 & -33 & 3.22e$^{-2}$ & 322 & -2.424 & 277  \\
\textbf{Pa 5-3}  & \textbf{1281.808} & \textbf{-67} & ... & ...  & \textbf{-11.4}  & \textbf{340}  \\
Br 25-4 & 1496.733 & -43 & 4.70e$^{-2}$ & 295 & -0.289 & ...  \\
Br 24-4 & 1500.086 & -37 & 4.38e$^{-2}$ & 301 & -0.275 & ...  \\
Br 23-4 & 1503.904 & -48 & 5.56e$^{-2}$ & 318 & -0.360 & ...  \\
Br 22-4 & 1508.277 & -81 & 8.26e$^{-2}$ & 203 & -0.338 & 270  \\
Br 21-4 & 1513.322 & -77 & 7.10e$^{-2}$ & 380 & -0.537 & 198  \\
Br 20-4 & 1519.184 & -37 & 7.61e$^{-2}$ & 281 & -0.443 & ...  \\
Br 19-4 & 1526.054 & -84 & 1.03e$^{-1}$ & 211 & -0.426 & ...  \\
Br 18-4 & 1534.179 & -51 & 1.08e$^{-1}$ & 297 & -0.632 & ...  \\
Br 17-4 & 1543.892 & -54 & 1.21e$^{-1}$ & 330 & -0.786 & 258  \\
Br 16-4 & 1555.645 & -84 & 1.57e$^{-1}$ & 410 & -1.161 & 189  \\
Br 15-4 & 1570.066 & -47 & 1.72e$^{-1}$ & 333 & -1.050 & 248  \\
Br 14-4 & 1588.054 & -48 & 2.24e$^{-1}$ & 331 & -1.386 & 242  \\
Br 13-4 & 1610.931 & -49 & 2.60e$^{-1}$ & 323 & -1.524 & 244  \\
Pa 4-3  & 1875.101 & -61 & 1.01e$^{1}$  & 343 & -40.337 & ...  \\
Br 8-4  & 1944.556 & -48 & 1.12         & 326 & -4.167  & 304  \\
\textbf{Br 7-4}  & \textbf{2165.529} & \textbf{-73} & ...   & ... & \textbf{-7.62}  & \textbf{373}  \\
Pf 25-5 & 2373.729 & -61 & 9.96e$^{-2}$ & 407 & -0.479 & ...  \\
Pf 24-5 & 2382.173 & -41 & 1.27e$^{-1}$ & 380 & -0.577 & ...  \\
Pf 23-5 & 2391.815 & -65 & 1.77e$^{-1}$ & 308 & -0.643 & ...  \\
Pf 21-5 & 2415.726 & -57 & 2.02e$^{-1}$ & 339 & -0.737 & ...  \\
Pf 20-5 & 2430.699 & -54 & 2.37e$^{-1}$ & 324 & -0.920 & ...  \\
Pf 19-5 & 2448.332 & -49 & 2.82e$^{-1}$ & 333 & -1.133 & ...  \\
\hline\\[-1.5ex]
\end{tabular}
\end{minipage}
\begin{minipage}[t]{0.5\textwidth}
\centering
\vspace{0pt}
\centering
\caption{Identified \ion{He}{1} lines\label{tab:line:helium}}
\begin{tabular}{lrlllll}
\hline\hline\\[-1.5ex]
$\lambda_0$ & V$_{h}$ & Flux & \textsc{fwhm}       & \textsc{eqw}  & V$_{PC}$ \\
\hline\\[-1.5ex]
 706.518   & -123 & 7.75e$^{-5}$ & 342 & -1.820 & 115\\
 728.135   & -127 & 5.30e$^{-5}$ & 327 & -0.661 & 133\\
1031.122  & -118 & 6.86e$^{-3}$ & 450 & -0.633 & 147\\
1196.904  & -116 & 2.34e$^{-2}$ & 437 & -0.608 & 179\\
1252.751  & -73  & 3.28e$^{-2}$ & 223 & -0.315 & ...  \\
1279.050  & -118 & 6.36e$^{-2}$ & 373 & -0.326 & 316\\
1278.491  &  14  & 6.51e$^{-2}$ & 386 & -0.923 & 312\\
1296.843  & -77  & 2.82e$^{-2}$ & 203 & -0.211 & 193\\
1700.234  & -111 & 2.49e$^{-1}$ & 278 & -1.100 & 164\\
2058.129  & -133 & 1.81         & 299 & -6.120 & 135\\
2112.002  & -119 & 9.89e$^{-2}$ & 296 & -0.471 & 360\\
2161.701  & -111 & 1.01e$^{-1}$ & 426 & -0.164 & 280\\
\hline\\[-1.5ex]
\end{tabular}

\vspace{0pt}
\smallbreak
\caption{Identified transitions of \ion{Mg}{2}\label{tab:line:mg}}
\begin{tabular}{lllllll}
\hline\hline\\[-1.5ex]
Term & $\lambda_0$ & V$_{h}$ & Flux & \textsc{fwhm}    & \textsc{eqw} \\ 
\hline\\[-1.5ex]
2P$_0$-2D &  787.705 & -94.5  & 2.78e$^{-4}$ & 413 & -0.514 \\
2P$_0$-2D &  789.637 & -59.4  & 4.62e$^{-4}$ & 329 & -0.708 \\
2P$_0$-2S &  821.398 & -10.3  & 4.39e$^{-4}$ & 236 & -0.206 \\
2P$_0$-2S &  823.464 & -117.9 & 5.84e$^{-4}$ & 355 & -0.411 \\
2S-2P$_0$ &  921.825 & -125.3 & 1.02e$^{-2}$ & 270 & -0.988 \\
2S-2P$_0$ &  924.426 & -52.3  & 8.3e$^{-3}$  & 325 & -0.980 \\
2D-2P$_0$ & 1091.424 & -58.0  & 4.5e$^{-2}$  & 283 & -1.484 \\
2D-2P$_0$ & 1095.177 &  17.9  & 2.0e$^{-2}$  & 273 & -0.589 \\
2S-2P$_0$ & 2136.90  & -97.4  & 2.37e$^{-1}$ & 351 & -0.948 \\
2S-2P$_0$ & 2143.22  & -82.4  & 1.27e$^{-1}$ & 350 & -0.509 \\
2D-2P$_0$ & 2412.46  & -110.1 & 1.31e$^{-1}$ & 344 & -0.534 \\
2D-2P$_0$ & 2404.15  & -154.9 & 3.82e$^{-1}$ & 353 & -1.600 \\
\hline\\[-1.5ex]
\end{tabular}

\vspace{0pt}
\smallbreak
\caption{Optical forbidden transitions\label{tab:line:forbidden}}
\begin{tabular}{lllllll}
\hline\hline\\[-1.5ex]
         & Term & $\lambda_0$ & V$_{h}$ & Flux & \textsc{fwhm} & \textsc{eqw}  \\
\hline\\[-1.5ex]
$[$\ion{O}{1}$]$  & 3P-1D   & 630.030 & -32 & 1.0e$^{-4}$ & 11 & -2.21 \\
$[$\ion{N}{2}$]$ & 3P-1D   & 654.804 & -30 & 1.6e$^{-4}$ & 17 & -0.72 \\
$[$\ion{N}{2}$]$ & 3P-1D   & 658.346 & -32 & 5.1e$^{-4}$ & 6  & -2.09 \\
$[$\ion{S}{2}$]$ & 4S$_0$-2D$_0$ & 671.644 & -29 & 3.4e$^{-4}$ & 15 & -1.46 \\
$[$\ion{S}{2}$]$ & 4S$_0$-2D$_0$ & 673.081 & -31 & 1.9e$^{-4}$ & 8  & -0.75 \\
\hline\\[-1.5ex]
\end{tabular}

\end{minipage}
\end{table}

\begin{table}[h]
\footnotesize
\caption{Identified iron transitions (FW: HWHM of the flat profile, GW: HWHM of the gaussian broadening, both in km/s)\label{tab:line:iron}}
\centering
\begin{tabular}{llrlrlll}
\hline\hline\\[-1.5ex]
         & Term & $\lambda$  & V$_{helio}$ & Flux (mJy)  & EQW  & FW          & GW          \\
\hline\\[-1.5ex]
\ion{Fe}{2}   & z4F$_0$-b4G      &  995.631 & -79\p 3 & 3.4\p 0.05    & -0.254 & 220\p 4 & ~~94\p 8 \\
\ion{Fe}{2}   & z4F$_0$-b4G      &  999.758 & -63\p 2 & 434.0\p 0.17  & -2.914 & 217\p 2 & 117\p 3  \\
\ion{Fe}{2}   & z4D$_0$-d2F      & 1017.392 & -83\p 7 & 6.1\p 0.22    & -0.647 & 252\p 7 & 104\p 18 \\
\ion{Fe}{2}   & z4F$_0$-d2F      & 1043.476 & -75\p 3 & 3.7\p 0.08    & -0.332 & 265\p 3 & ~~69\p 8 \\
\ion{Fe}{2}   & z4F$_0$-b4G      & 1050.150 & -82\p 4 & 26.0\p 0.42   & -0.047 & 236\p 4 & ~~97\p 9 \\
\ion{Fe}{2}   & z4F$_0$-b4G      & 1112.558 & -72\p 2 & 38.1\p 0.37   & -1.811 & 233\p 2 & ~~93\p 5 \\
$[$\ion{Fe}{2}$]$ & a6D-a4D   & 1256.680 & -47\p 3 & 11.7\p 0.2~~  & -0.243 & 295\p 4 & ~~65\p 8 \\	
\ion{Fe}{2}   & z4F$_0$-c3F      & 1687.320 & -73\p 2 & 495.0\p 1.7~~ & -3.612 & 257\p 2 & ~~66\p 4 \\
\ion{Fe}{2}   & z4D$_0$-c4F      & 1741.401 & -88\p 1 & 152.9\p 2.0~~ & -1.026 & 282\p 6 & 138\p 9  \\
\ion{Fe}{2}   & z4F$_0$-c3F      & 1974.611 & -74\p 3 & 282.1\p 4.7~~ & -1.598 & 266\p 3 & ~~80\p 7 \\
\ion{Fe}{2}   & z4D$_0$-c4F      & 1986.841 & -76\p 3 & 81.1\p 1.1~~  & -0.484 & 265\p 3 & ~~80\p 6 \\
$[$\ion{Fe}{2}$]$  & a4P-a2P  & 2046.007 & -55\p 7 & 24.0\p 0.64   & -0.139 & 280\p 7 & ~~55\p 11\\
\ion{Fe}{2}   & z4F$_0$-c3F      & 2088.810 & -76\p 2 & 299.0\p 1.6~~ & -1.690 & 257\p 2 & ~~81\p 4 \\
$[$\ion{Fe}{2}$]$  & a2G-a2H  & 2223.760 & -40\p 6 & 17.9\p 0.6~~  & -0.096 & 286\p 6 & ~~52\p 15\\
\ion{Fe}{2}   & z4D$_{3/2}$-c4P$_{3/2}$ & 2240.152  & -62\p 3 & 28.4\p 0.5~~  & -0.140 & 274\p 3 & ~~54\p 7 \\
\hline\\[-1.5ex]
\end{tabular}
\end{table}

\bibliographystyle{aasjournal}
\bibliography{./references.bib}

\end{document}